\documentclass[aps,prb,twocolumn,superscriptaddress,showpacs]{revtex4-1}
\usepackage{graphicx}
\usepackage{graphics}
\usepackage{amsmath}
\usepackage{amsfonts}
\usepackage{amssymb}
\usepackage{ifpdf}
\usepackage{epstopdf}
\usepackage{makeidx}
\usepackage{epsfig}
\usepackage{bm}
\usepackage{color}
\usepackage{xcolor}
\usepackage[unicode=true, bookmarks=false,breaklinks=false,pdfborder={0 0 1},backref=false,colorlinks=false]{hyperref}
\hypersetup{hypertex}
\usepackage{breakurl}
\hypersetup{backref,colorlinks=true,citecolor=blue,linkcolor=blue,filecolor=blue,urlcolor=blue}

\begin{document}

\newcommand{\beq}   {\begin{equation}}
\newcommand{\eeq}   {\end{equation}}
\newcommand{\bea}   {\begin{eqnarray}}
\newcommand{\eea}   {\end{eqnarray}}
\newcommand{\ba}   {\begin{align}}
\newcommand{\ea}   {\end{align}   }
\newcommand{\ve}   {\varepsilon}
\newcommand{\x} {\mathbf{x}}
\newcommand{\X} {\mathbf{X}}
\newcommand{\nn} {\nonumber}
\newcommand{\CAIO}[1]{\textcolor{red}{\fbox{Caio} {\sl#1}}}
\newcommand{\ENRIQUE}[1]{\textcolor{blue}{\fbox{Enrique} {\sl#1}}}
\newcommand{\Alexis}[1]{\textcolor[rgb]{1,0,0.5}{\fbox{El Chechi} {#1}}}
\title{Gauge fields in graphene with nonuniform elastic deformations: 
\\A quantum field theory approach}

\author{Enrique Arias}
\affiliation{Instituto Polit\'ecnico,
  Universidade do Estado do Rio de Janeiro, 28625-570 Nova Friburgo, Brazil}
\author{Alexis R. Hern\'andez}
\affiliation{Instituto de F\'{\i}sica, 
Universidade Federal do Rio de Janeiro, 21941-972 Rio de Janeiro, Brazil}
\author{Caio Lewenkopf}
\affiliation{Instituto de F\'{\i}sica,
  Universidade Federal Fluminense, 24210-346 Niter\'oi, Brazil}

\begin{abstract}
We investigate  the low energy continuum limit theory for electrons in a graphene 
sheet under strain. We use the quantum field theory in curved spaces to 
analyze the effect of the system deformations into an effective gauge field. 
We study both in-plane and out-of-plane deformations 
and obtain a closed expression for the effective gauge field due to arbitrary nonuniform 
sheet deformations. The obtained results reveal a remarkable relation between the local-pseudo 
magnetic field and the Riemann curvature, so far overlooked. 

\end{abstract}
\pacs{73.22.Pr,04.62.+v}
\maketitle

\section{Introduction} 
\label{sec:purpose}

One of the most active research topics in graphene, as well as in other 2D materials,
is the study of the interplay between their electronic and mechanical 
properties \cite{Bunch2007,Garcia2008,Chen2009,Eichler2011,Jones2014}. 
Among the most remarkable results the possibility of using strain to generate 
large effective magnetic fields, with considerable effects on the electronic dynamics, 
has attracted a lot of attention \cite{Guinea2009,Pereira09b,Low2011}. 
These results triggered several interesting strain engineering proposals,
such as the use of strain for quantum electronic pumping \cite{Prada2009,Low12}, 
generation of pure bulk valley currents \cite{Jiang2013}, and for 
confining electrons \cite{Zhu2014,Bastos2014}, to name a few.

There are two main theoretical approaches that cast strain/deformation 
induced modifications in the electronic properties of graphene monolayers in terms 
of effective gauge fields. The most standard one 
\cite{Suzuura2002,Manes2007,Guinea08,Manes2013,deJuan13,Masir2013}  is based 
on the low-energy continuum limit of a tight-binding model that accounts for
the displacements of the carbon atoms in a strained graphene sheet.
This approach has been successful in explaining the local density of states inferred 
by scanning probe spectroscopy  experiments in graphene nanobubbles 
\cite{Levy2010,Yeh2011}. In contrast, to be consistent with transport 
experiments \cite{Lundeberg10,Burgos2015}, the pseudo-magnetic fields have to be
renormalized \cite{Suzuura2002,Midtvedt2015}. 
Recent studies have further developed the theory showing the modifications in the
effective theory due to the non-Bravais nature of the graphene primitive unit cell 
\cite{Midtvedt2015} and the effects of deformations in the structure of the 
reciprocal space \cite{OlivaLeyva2013}.

The second approach is based on the quantum field theory in curved spaces 
\cite{Weinberg1972,Birrell1982}. It starts with the low-energy effective Dirac 
equation for graphene \cite{Semenoff84} and, by considering a curved space 
metric, obtains geometry-induced gauge fields \cite{Vozmediano10,deJuan07}. 
This theory was the first to predict a space-dependent Fermi velocity, that has been 
experimentally confirmed  \cite{Yan2013}.
It has been recently shown \cite{Sanjuan2014a} that this approach can be extended 
beyond the continuum limit by using discrete differential geometry. 
The effective theory in curve spaces has also fostered interesting research 
in materials other than graphene \cite{Khveshchenko2013,Volovik2014}.

The current theoretical status \cite{deJuan13,Manes2013} is that there is no unified 
approach that combines elasticity theory with the continuum limit of the corresponding 
tight-binding approximation \cite{Guinea08} and the quantum field theory in curved geometry 
(or geometrical approach) \cite{deJuan07,Vozmediano10}. 
Symmetry arguments \cite{Manes2013} indicate that there is room for improving both 
approaches.
With this motivation our study focuses on further developing the geometric approach.

In this paper we advance the quantum field theory approach put forward in the seminal 
works of Vozmediano and collaborators \cite{deJuan07,Vozmediano10}, that addressed 
a case with a simple geometry, namely, a Gaussian bump deformation.
We derive a general expression for the pseudo-magnetic field for an arbitrary nonuniform 
graphene surface deformation. By restricting our analysis 
to the case of out-of-plane lattice deformations, we are able to express the pseudo-magnetic 
field in a simple analytical form. This result reveals a remarkable relation between the 
pseudo-magnetic field and the Riemann curvature.

We further generalize these findings for the realistic case that combines in-plane and out-of-plane 
deformations by incorporating elements of elasticity theory \cite{Landau1986}. 
Here it is still possible to solve the problem and we can numerically verify that the relation between
the pseudo-magnetic field and the scalar curvature holds. Unfortunately, the expressions for 
the gauge field induced by an arbitrary strain become very lengthy and not particularly insightful.
For this reason, we use the simple Gaussian bump deformation to illustrate our results. 

This paper is organized as follows. In Sec.~\ref{sec:Dirac} we review the theory of Dirac fermions 
in flat and curved spaces. In Sec.~\ref{sec:model} we use this formalism to model the 
wave equation of quasi particles in rippled graphene. Next, we introduce a parameterization for a 
surface with an arbitrary curvature. By correctly defining the pseudo-magnetic field, we find a 
connection between the graphene curvature and the induced pseudo-magnetic field.
In Sec.~\ref{sec:critique} we discuss the local plane frame associated to a given point in the curved 
surface that has been used by some authors \cite{deJuan12,Chaves14} and show that the identification 
of the gauge field in this position-dependent frame leads to an incorrect assessment of the 
pseudo-magnetic field.
We present our conclusions in Sec.~\ref{sec:conclusions}.

%
\section{Dirac equation in curved space} 
\label{sec:Dirac}

In this Section we briefly review the formalism of the Dirac equation in flat and 
curved spacetimes, paying particular attention to the technical issues most directly 
related to our study.  

The flat spacetime points are denoted by $X=(T,\X)$ whose components have \textit{flat} indices 
denoted by the greek letters $\{\alpha,\beta,\delta,...\}$. The Dirac equation in the flat Minkowski 
spacetime is written as \cite{Birrell1982}
\beq
(i\gamma^\alpha\partial_\alpha+m)\psi(X)=0,
\label{eq:Dirac_flat}
\eeq
where $\gamma^\alpha$ are the usual Dirac matrices and  $\partial_\alpha=
\partial/\partial X^\alpha$. Since we work with differentiable functions the derivatives 
do commute, $[\partial_\alpha,\partial_\beta]=0$.

In the presence of a gravitational field, Eq.~\eqref{eq:Dirac_flat} needs to be modified 
to account for a spacetime structure. 
In this case, the curved spacetime points are denoted by $\chi$ whose \textit{curved} 
components we designate by the greek letters $\{\mu,\nu,\kappa,...\}$.
The standard procedure for tensor (bosonic) fields is to realize the minimal coupling 
of the field with gravity by substituting the Minkowski spacetime metric tensor 
$\eta_{\alpha\beta}$ by the general Riemannian metric $g_{\mu\nu}$ and by the replacing 
of the usual derivative $\partial_\alpha$ by the covariant derivative 
$\nabla_\mu$. 
For spinorial (fermionic) fields this procedure is inadequate, due to the lack of a spinorial 
covariant derivative in terms of the metric.
To correctly account for the coupling of the spinor with the curved spacetime one uses the 
vierbeins formalism \cite{Weinberg1972,Birrell1982}.

The equivalence principle, that is the basis of the geometric theory of gravitation, states that 
one can not locally distinguish between a real gravitational field and the effects caused by a 
non-inertial reference frame \cite{Ramond1990}. 
This implies that in the neighborhood $\cal{X}$ of any given point $\chi_p$ in curved spacetime, 
one can find a local reference frame $\{\xi^a_{\cal X}\}$ such that all the effects of gravity vanish.

\begin{figure}[h]
\begin{center}
\centering \includegraphics[width=0.7\columnwidth]{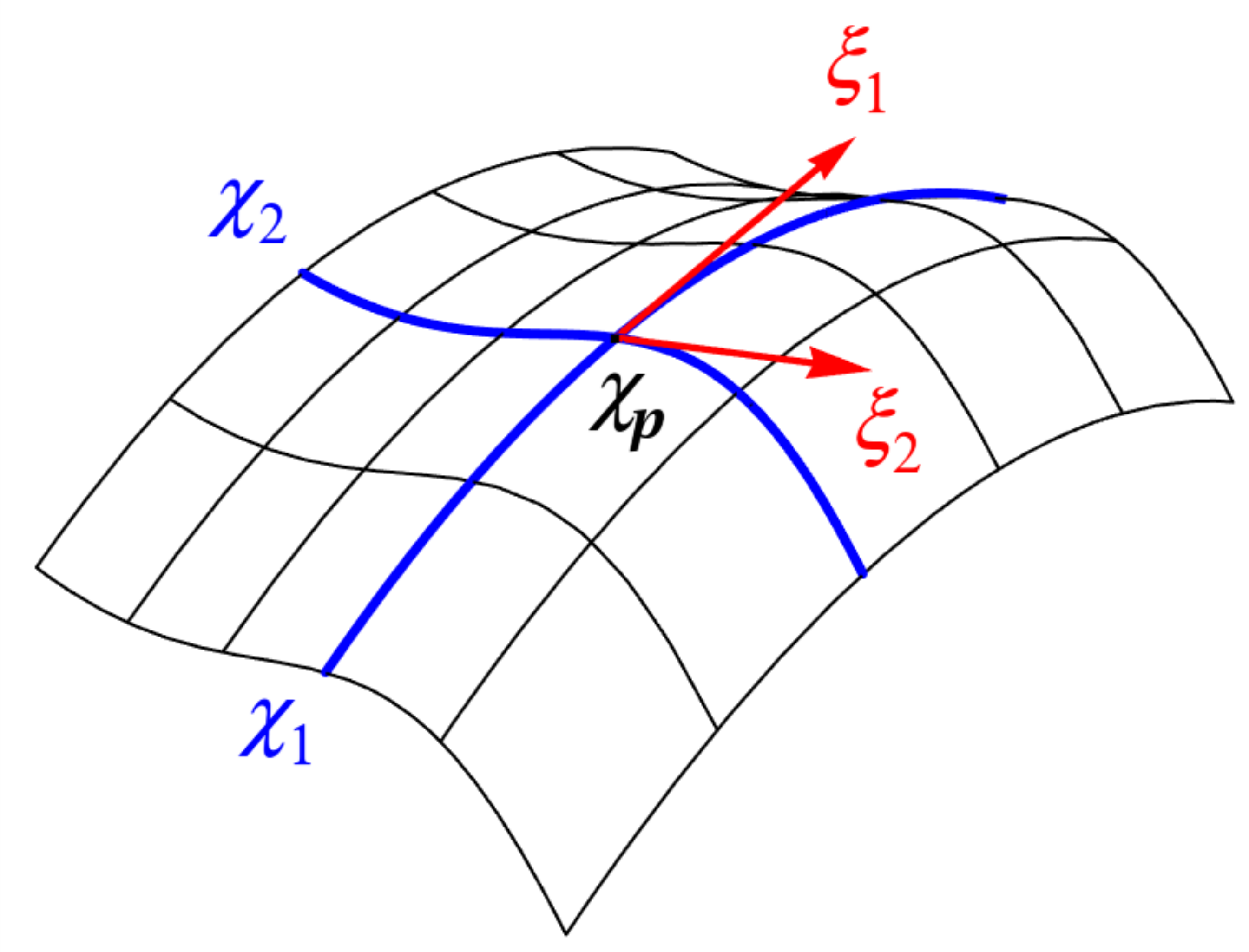} 
\caption{Illustration of a curved surface: $(\chi_1,\chi_2)$ correspond to global-curved coordinates 
and $(\xi_1,\xi_2)$ to local-flat coordinates taken at a given point $\chi_p$.}
\label{fig:coordinates}
\end{center}
\end{figure}

This local-flat space has the Minkowski metric $\eta_{ab}$ and the distance between infinitesimally 
close points is $ds^2=\eta_{ab}\,d\xi^a_{\cal X}\,d\xi^b_{\cal X}$. Near the point $\chi_p$ the same 
distance, given in terms of the curved coordinates, is $ds^2=g_{\mu\nu}d\chi^\mu d\chi^\nu$. 
By comparing these expressions and considering that $\xi^a_{\cal X}$ 
are local functions of $\chi_p$, one concludes that
\beq
g_{\mu\nu}({\chi_p})=\eta_{ab}\,e_\mu^{\,\,\,a}({\chi_p})e_\nu^{\,\,\,b}(\chi_p),
\eeq 
where $e_\mu^{\,\,\,a}({\chi_p})$ are the so-called vierbeins, defined as 
\cite{Weinberg1972,Birrell1982}
\beq
e_\mu^{\,\,\,a}(\chi_p)=\frac{\partial\xi^a_{\cal X}(\chi)}{\partial\chi^\mu}\Big|_{\chi=\chi_p}.
\eeq 
The inverse vierbeins read $e^\mu_{\,\,\,a}=g^{\mu\nu}\eta_{ab}\,e_\nu^{\,\,\,b}$ and 
satisfy $g^{\mu\nu}=\eta^{ab}\,e^\mu_{\,\,\,a}e^\nu_{\,\,\,b}$.
Note that these definitions are local. 

In summary, in this paper we define three coordinate systems: 
(i) The global-flat Minkowski space with coordinates $X$, and component
indices $(\alpha, \beta, ...)$; 
(ii) The global-curved space with coordinates $\chi$ labeled by $(\mu, \nu, ...)$; 
(iii) The local-flat position dependent reference system with coordinates $\xi$ and 
local-flat indices $(a, b, ...)$.

The vierbeins are operators that transform quantities in the global-curved 
space to their corresponding local-flat counterparts. For example, one can find the curved 
version of the Dirac matrices, $\gamma^\mu$, by contracting the flat gamma matrices 
$\gamma^a$ with the vierbeins, namely
\beq
\gamma^\mu=e^\mu_{\,\,\,a}\gamma^a.
\eeq 
It can be verified that these matrices satisfy the generalized Clifford algebra in the 
curved spacetime $\{\gamma^\mu,\gamma^\nu\}=2g^{\mu\nu}$. Also, the covariant 
derivative $\nabla_\mu$ of a spinorial field can now be properly defined as
\beq
\label{eq:covariant_derivative}
\nabla_\mu\psi=(\partial_\mu+\Omega_\mu)\psi,
\eeq
where $\Omega_\mu$ is the so called spin connection, defined by
\beq
\label{eq:spin-con}
\Omega_\mu=\frac{1}{2}\omega_{\mu ab}\,\Sigma^{ab},
\eeq
with
\beq
\omega_{\mu ab}=e^\nu_{\,\,a}\nabla_\mu e_{\nu b}-e^\nu_{\,\,\ b}\nabla_\mu e_{\nu a}.
\label{eq:spin-con-comp}
\eeq
The covariant derivatives of the vierbeins read
$\nabla_\mu e_{\nu a}=\partial_\mu e_{\nu a}-\Gamma_{\mu\nu}^{\,\,\,\ \kappa} e_{\kappa a}$ 
and the affine connections  $\Gamma_{\mu\nu}^{\,\,\,\,\,\kappa}$ 
are the usual Christoffel symbols \cite{Weinberg1972}.
The operators $\Sigma^{\alpha\beta}$ are the generators of the Lorentz group in the 
spinorial space, namely, $\Sigma^{\alpha\beta}=[\gamma^\alpha,\gamma^\beta]/4$. 

Hence, the Dirac equation in a curved spacetime is given by
\beq
\left(i\gamma^\mu\nabla_\mu+m\right)\psi(\chi)=0.
\eeq
The covariant derivative, given by Eq.~\eqref{eq:covariant_derivative}, allows 
one to interpret $\Omega_\mu$ as an effective gauge field induced by the 
space curvature \cite{deJuan07,Vozmediano10}. The analysis of the
pseudo-magnetic fields associated to $\Omega_\mu$ is the main focus of this 
study. 

We note that the space dependence of $\gamma^\mu$ gives rise to another 
important effect \cite{deJuan07,Vozmediano10}, namely, the renormalization of the 
electron velocity, called Fermi velocity $v_F$ in the graphene literature. 
Our analysis reproduces the same effects showed 
Refs.~\onlinecite{deJuan07,Vozmediano10} for the spacial dependence of $v_F$.

In summary, by knowing all the geometric objects associated to a given 
spacetime, one can establish the influence of the space curvature on the dynamics of 
the particles in that space. 
In the following Section we describe how to find the general  geometric properties of 
an arbitrary curved graphene sheet. We then identify and discuss the correct 
pseudo-magnetic field generated by the corrugations.

%
\section{Application to curved graphene} 
\label{sec:model}

We now adapt the results of Sec. \ref{sec:Dirac} to describe the dynamics of electrons 
in rippled graphene by coupling a quenched curved background to the quasi-particle wave 
function. The procedure we present follows the ideas put forward by Vozmediano and 
collaborators \cite{Vozmediano10} and generalizes their results to arbitrary sheet profiles. 
To that end, we have found a way to properly treat the gauge invariance associated to 
arbitrary curved spaces. This allowed us to identify a remarkable relation between the 
pseudo-magnetic field and the Riemann curvature. This relation implies that our results are 
manifestly gauge invariant, in distinction to previous studies. 

Following the outline presented in Sec \ref{sec:Dirac}, we describe the dynamics
of electrons constrained to propagate in a two-dimensional curved surface by finding the metric, 
the vierbeins, the affine connection, and the spin connection associated to it.

\subsection{Out-of-plane deformations} 
\label{sec:out-plane}

The geometry of a rippled two-dimensional surface can be parametrized by the 
function $z=h(x,y)$, where $h$ represents the out of plane deviation from the 
flat surface $z=0$.
In the notation introduced in the previous section, $h(\chi)$ is the surface height
at the position $\chi=(x,y)$. Since the coordinates $(x,y)$ parameterize the curved 
surface, in general the intersection of the plane defined by a constant $x$ (or $y$) with 
the surface $h(x,y)$ is not a straight line. Further, these curvilinear coordinates do 
not need to be orthogonal on the surface. 

The square distance between two infinitesimally close points on the curved 
surface is given by
\begin{align}
\label{e1}
ds^2= & \left[1+(\partial_xh)^2\right]dx^2+\left[1+(\partial_yh)^2\right]dy^2+
\nonumber\\ & 
2(\partial_xh)(\partial_yh)dxdy.
\end{align}
We write Eq.~\eqref{e1} as $ds^2=g_{\mu\nu}d\chi^\mu d\chi^\nu$.
For notation convenience, to establish a clear difference between curved and flat 
indices, we call $(\chi^x, \chi^y)$ \textit{curved coordinates}, whose ``curved'' indices 
are denoted by $\mu=\{x,y\}$. 
In terms of the curved coordinates the metric on the manifold reads
\beq
g_{\mu\nu}=\left(
\begin{array}{cc}
 1+h_x^2 & h_xh_y\\
 h_xh_y  & 1+h_y^2
 \label{e2}
\end{array}
\right),
\eeq
where we introduce the shorthand notation $h_\mu=\partial_\mu h$. 
It is convenient, for later use, to write $g_{\mu\nu}$ in a compact form as
\beq
g_{\mu\nu}=\delta_{\mu\nu}+h_\mu h_\nu.
\label{10}
\eeq

As already discussed, locally at each point $\chi$ of the curved 
surface, the manifold can be seen as flat, and there exits a local 
coordinate system such that in a small region ${\cal X}$, near the 
point $\chi$, the metric is Euclidean.
We denote this local flat coordinates by $\xi_{\cal X}^a=\xi^a_{\cal X}(\chi)$, 
with ``local-flat'' indices $a=\{1,2\}$. 
In terms of the local-flat coordinates, the square distance between two points 
is given by $ds^2=\delta_{ab}\,\,d\xi^a_{\cal X}\, d\xi^b_{\cal X}$. Comparing 
this expression with $ds^2=g_{\mu\nu}d\chi^\mu d\chi^\nu$ defined above,
we find the relation between both metrics, namely
\beq
g_{\mu\nu}(\chi)=\delta_{ab}\,e_\mu^{\,\,\,a}(\chi)\,e_\nu^{\,\,\,b}(\chi).
\label{e3}
\eeq
Due to the symmetry $g_{\mu\nu}=g_{\nu\mu}$, Eq.~(\ref{e3}) gives three 
independent relations to find the vectors $e_\mu^{\,\,\,1}=(e^{\,\,\,1}_x,e^{\,\,\,1}_y)$ 
and $e_\mu^{\,\,\,2}=(e^{\,\,\,2}_x,e^{\,\,\,2}_y)$.
In this way, except for a single degree of freedom, the vierbeins are determined 
by the metric.
It will be shown that this indetermination does not manifest
in physical observables.
By simple algebraic manipulation we obtain a general solution for the vierbeins, 
namely
\beq
e_\mu^{\,\,\,a}=\left(
\begin{array}{cc}
 \sqrt{1+h_x^2}\cos\theta & \sqrt{1+h_x^2}\sin\theta             
 \\
 \sqrt{1+h_y^2}\cos\bar\theta & \sqrt{1+h_y^2}\sin\bar\theta
\end{array}
\right),
\label{e5}
\eeq
where  $\theta=\theta(x,y)$  is an arbitrary function related to the orientation 
of the local-flat coordinate axis at the point $\chi=(x,y)$. 
In Eq. (\ref{e5}) we also introduced $\bar{\theta}$, given by
\beq
\bar{\theta}=\theta+\arccos\left(\frac{h_{x}h_{y}}{\sqrt{(1+h_y^2)(1+h_y^2)}}\right).
\label{e6}
\eeq
Let us now calculate the inverse vierbeins, 
$e^\mu_{\,\,\,a}=g^{\mu\nu}\delta_{ab}e_\nu^{\,\,\,b}$.
From Eq. (\ref{e2}) we obtain the inverse metric
\beq
g^{\mu\nu}=\frac{1}{1+h_x^2+h_y^2}\left(
\begin{array}{cc}
 1+h_y^2 & -h_xh_y \\
 -h_xh_y & 1+h_x^2
\end{array}
\right).
\label{e8}
\eeq
or
\beq
g^{\mu\nu}=(\delta^{\mu\nu}+h^*_\mu h^*_\nu)/(1+h_x^2+h_y^2),
\label{11}
\eeq
where we have introduced the dual $h^*_\mu=\varepsilon_{\mu\nu}h_\nu$, 
being $\varepsilon_{\mu\nu}$ the Levi-Civita tensor in two dimensions, {\it i.e.},
 $\varepsilon_{12}=-\varepsilon_{21}=1$ and $\varepsilon_{11}=\varepsilon_{22}=0$.
By using Eqs. (\ref{e5}) and (\ref{e8}) we calculate the inverse vierbeins  
\begin{align}
e^\mu_{\,\,\,a}=\frac{1}{\sqrt{1+h_x^2+h_y^2}}\left(
\begin{array}{ll}
\sqrt{1+h_y^2}\sin\bar{\theta} & -\sqrt{1+h_y^2}\cos\bar{\theta}\\
-\sqrt{1+h_x^2}\sin\theta & \sqrt{1+h_x^2}\cos\theta
\end{array}
\right).
\label{9}
\end{align}

Having defined the metric tensor and the vierbeins, 
we proceed to calculate the spin connection that relates the wave function 
of the Dirac field with the curved space geometry. 
First, we determine the vierbeins covariant derivatives which are expressed
in terms of the affine connection. 
The affine connections coincide with the Christoffel's symbols in the case where 
the spacetime has a curvature but not a torsion \cite{Sabbata1985}.
Effects of torsion have been studied in the context of topological defects \cite{Vozmediano09},
which are not within the scope of this paper. Hence, 
\beq
\Gamma_{\mu\nu}^{\,\,\,\,\,\kappa}=\frac{1}{2}g^{\kappa\sigma}\left(\partial_\mu g_{\nu\sigma}
+\partial_\nu g_{\mu\sigma}-\partial_\sigma g_{\mu\nu}\right).
\label{12}
\eeq
By using Eqs. (\ref{10}) and (\ref{11}) we express the affine connection in terms of $h$, 
namely
\beq
\Gamma_{\mu\nu}^{\,\,\,\,\,\kappa}=\frac{h_{\mu\nu}\,\,h_\kappa}{1+h_x^2+h_y^2},
\label{13}
\eeq
where $h_{\mu\nu}=\partial^2_{\mu\nu} h$. 

Before we obtain the spin connection, it is instructive to present
the curvature tensors associated to the surface. 
The Riemann curvature tensor is given by
\beq
R_{\mu\nu\rho}^{\,\,\,\,\,\,\,\,\,\,\kappa}=\partial_\mu
\Gamma_{\nu\rho}^{\,\,\,\,\,\kappa}-\partial_\nu\Gamma_{\mu\rho}^{\,\,\,\,\,\kappa}
+\Gamma_{\mu\sigma}^{\,\,\,\,\,\kappa}\Gamma_{\nu\rho}^{\,\,\,\,\,\sigma}
-\Gamma_{\nu\sigma}c\Gamma_{\mu\rho}^{\,\,\,\,\,\sigma},
\eeq 
while the scalar curvature of the surface reads
\beq
{\cal R}=R_{\kappa\mu\nu}^{\,\,\,\,\,\,\,\,\,\,\kappa}g^{\mu\nu}.
\eeq 
Hence, for the two-dimensional graphene surface with out-of-plane deformations 
the scalar curvature reads
\beq
{\cal R}=2\,\frac{h_{xx}h_{yy}-h_{xy}^2}{(1+h_x^2+h_y^2)^2}.
\label{R}
\eeq 

Let us now resume the calculation of the spin connection $\Omega_\mu$ of the Dirac 
field for a general two-dimensional curved surface, defined by Eq.~\eqref{eq:spin-con} 
in terms of the components $\omega_{\mu ab}$ \cite{Weinberg1972}, given by  
Eq.~\eqref{eq:spin-con-comp}.
Due to the antisymmetry of the $\omega_{\mu ab}$ in its flat indices,
the only non-zero component is $\omega_{\mu 12}$.
One can identify the later with the effective gauge field induced by 
the curvature, and call ${\cal A}_\mu=\omega_{\mu 12}$ \cite{Vozmediano10}.

After some algebra, we obtain that, for a general two-dimensional curved surface 
described by the function $h=h(x,y)$, the effective gauge field is given by
\beq
{\cal A}_\mu=\frac{1}{\sqrt{1+h_x^2+h_y^2}}
\left(\frac{h_x\,h_{y\mu}}{1+h_y^2}-\frac{h_y\,h_{x\mu}}{1+h_x^2}\right) + 
\partial_\mu(\theta+\bar{\theta}).
\label{16}
\eeq
The arbitrary angle $\theta=\theta(x,y)$ enters as a total derivative 
in the gauge field ${\cal A}_\mu$ and therefore does not affect the 
pseudo-magnetic field  ${\cal B}=\mbox{rot}{\cal A}$, associated to ${\cal A}_\mu$. 
Arbitrary independent rotations of the local-flat frames along the curved surface, 
do not change the  pseudo-magnetic field and translate as the \textit{gauge 
transformations} of the field ${\cal A}_\mu$. 

The formal definition of the curl operator in non-orthogonal curvilinear coordinates is
\beq
{\cal B}=\frac{1}{\sqrt{g}}\,\varepsilon^{\mu\nu}\,\nabla_\mu{\cal A}_\nu,
\label{rot}
\eeq
where $g=det(g_{\mu\nu})$ and $\varepsilon^{\mu\nu}$ is the Levi-Civita symbol. 
By using the metric from Eq. (\ref{e2}) and the gauge field obtained in Eq.~\eqref{16}, 
we find that 
\beq
{\cal B}=2\,\frac{h_{xx}h_{yy}-h_{xy}^2}{(1+h_x^2+h_y^2)^2}.
\label{B}
\eeq
In two spacial dimensions the electromagnetic tensor becomes an anti-symmetric 
tensor of order three. This ensures that in two-dimensions the electric field is a vector, 
but the magnetic field has a single (scalar) component, perpendicular to the local-flat 
space associate to a given point in the deformed surface.
Hence, based on invariance arguments, one can expect ${\cal B}$ to be an
arbitrary function of the scalar curvature, namely, ${\cal B} = F({\cal R})$.
A direct comparison between Eqs.~(\ref{B}) and (\ref{R}) shows that the effective 
pseudo-magnetic field induced by ripples in graphene (in arbitrary units) is 
identical to the scalar curvature at each point of the surface 
\beq
{\cal B= R}.
\label{eq:identity}
\eeq 
The construction that leads to Eq.~\eqref{B} and the above identity are the main findings
of this paper.

We note that Eq.~\eqref{B} generalizes the results presented in Ref.~\onlinecite{deJuan07}, 
where the case of a radial symmetric $h(x,y)=f(r)$ was studied. For this simple geometry, 
the surface is conveniently parameterized by polar coordinates $(r,\phi)$. 
Here, our general expression for ${\cal A}_\mu$, Eq.~\eqref{16}, reduces to the one
found in Ref.~\onlinecite{deJuan07}. Using Eq. (\ref{B}), we find  
\beq
{\cal B}=\frac{2}{r}\frac{f'f''}{\left(1+f'^2\right)^2}.
\label{Bcorrect}
\eeq 
In distinction to our approach, Ref.~\onlinecite{deJuan07} obtains a pseudo-magnetic
field in the $z$-direction (perpendicular to the undeformed graphene sheet). 
There, the pseudo-magnetic field is defined by 
$B_z=(1/r)\partial_r(r{\cal A}_\phi)$, which gives $B_z=(2/r)f'f''(1+f'^2)^{-3/2}$. 
The discrepancy is due to the fact that Ref.~\onlinecite{deJuan07} calculates
$B_z$ using the curl operator in flat polar coordinates while ${\cal A}_\phi$ is 
expressed in curved coordinates.
Since the normal to the graphene surface is ${\bf n} = (-h_x, -h_y, 1)/(1+ h_x^2 
+ h_y^2)^{1/2}$, our pseudo-magnetic field projected in the $z$-direction is
${\cal B}/(1+f'^2)^{1/2}$.
Hence, the results are not consistent.
 
This difference prevents one to identify the pseudo-magnetic field with the 
scalar curvature, but is unlikely to impact on the analysis of experiments, where
the value of rms($f'$) is typically smaller than 0.1
\cite{Ishigami07,Geringer09,Burgos2015}.

\subsection{The general case: in-plane and out-of-plane deformations} 
\label{sec:in-and-out}

Here we analyze the general case of out-of-plane combined with in-place 
deformations. Now the infinitesimal distance between points at the corrugated 
surface is $ds^2=g_{\mu\nu}d\chi^\mu d\chi^\nu$, where the metric is 
expressed as
\beq
\label{eq:in-plane_g}
g_{\mu\nu}=\delta_{\mu\nu}+2u_{\mu\nu},
\eeq 
in terms of the deformation tensor, defined by the in-plane displacement 
vectors $u_\mu=u_\mu(x,y)$ and by the out-of-plane deformations $h(x,y)$ 
as follows
\beq
\label{eq:u_ab}
u_{\mu\nu}=\frac{1}{2}(\partial_\mu u_\nu+\partial_\nu u_\mu+2h_\mu h_\nu).
\eeq
Equations \eqref{eq:in-plane_g} and \eqref{eq:u_ab} are identical to those used in 
elasticity theory \cite{Landau1986}. 
This similarity will be explored in the next section to obtain material dependent 
expressions for the displacement vector components $u_{\mu}(x,y)$ in terms of the 
sheet topography $h(x,y)$.

Having defined the metric of the two-dimensional graphene surface we can proceed 
as before. The derivation has the same structure, the differences appear in the explicit 
expressions for the vierbeins, which are modified by the new physical ingredients. 
The general metric is
\beq
g_{\mu\nu}=\left(
\begin{array}{cc}
 g_{xx} & g_{xy} \\
 g_{xy} & g_{yy}\\
\end{array}
\right).
\label{}
\eeq
The solutions for the vierbeins read 
\beq
e_\mu^{\,\,\,a}=\left(
\begin{array}{cc}
 \sqrt{g_{xx}}\cos\theta & \sqrt{g_{xx}}\sin\theta \\
 \sqrt{g_{yy}}\cos\bar{\theta} & \sqrt{g_{yy}}\sin\bar{\theta}
\end{array}
\right).
\label{n}
\eeq
It can be checked that they satisfy $g_{\mu\nu}=\delta_{ab}e_\mu^{\,\,a}e_\nu^{\,\,b}$.
As before, in Eq. (\ref{n}), the variable $\theta=\theta(x,y)$ is an arbitrary angle 
related to the freedom of the orientation axes of local flat frames at point $\chi$. 
Here we define
\beq
\bar{\theta}=\theta+\arccos\left(\frac{g_{xy}}{\sqrt{g_{xx}g_{yy}}}\right).
\eeq 
To calculate the inverse vierbeins $e^\mu_{\,\,\,a}=g^{\mu\nu}\delta_{ab}e_\nu^{\,\,\,b}$, 
we need the inverse metric, namely
\beq
g^{\mu\nu}=\frac{1}{g}\left(
\begin{array}{rr}
 g_{yy} & -g_{xy} \\
 -g_{xy} & g_{xx}\\
\end{array}
\right).
\label{}
\eeq
As a result, the inverse vierbeins are
\beq
e^\mu_{\,\,\,a}=\frac{1}{\sqrt{g}}\left(
\begin{array}{rr}
\sqrt{g_{yy}}\sin\bar{\theta} & -\sqrt{g_{yy}}\cos\bar{\theta}\\
-\sqrt{g_{xx}}\sin\theta & \sqrt{g_{xx}}\cos\theta
\end{array}
\right).
\label{}
\eeq

Given ${\bf u}$ and $h$, these elements allow us to evaluate the effective 
pseudo-magnetic field, ${\cal B}$, generated by a combinations of 
out-of-plane and in-plane corrugations in graphene. 
However, in distinction to the out-of-plane case, the problem does not have 
a simple analytical solution.
We still verify that ${\cal B} = {\cal R}$ for a number of different models for 
strains and corrugations. 
In the forthcoming section we illustrate our results by analysing the case of 
a simple geometry $h(x,y)$.

\subsection{Application: Gaussian deformation in graphene}
\label{sec:application}

To construct the deformation tensor $u_{\nu\nu'}$, Eq.~\eqref{eq:u_ab}, 
it is necessary to know both $h(x,y)$ and the displacement vector fields 
$u_\nu(x,y)$.  The elasticity theory allows one to relate these quantities, 
as follows.
We consider the simplified scenario where we neglect shear forces between 
the substrate and the two-dimensional material under analysis. 
We follow the procedure suggested by Guinea and collaborators \cite{Guinea08}, 
namely, we assume that the system has a given system topography $h(x,y)$ 
and minimize the elastic energy by varying the in-plane degrees of freedom to 
obtain ${\bf u}(x,y)$.

The Hamiltonian corresponding to the elastic degrees of freedom reads \cite{Guinea08}
\beq
{\cal H}_{\rm elastic}=\!\int \! d{\bf r} \left\{ \frac{\lambda}{2}\left[\sum_\nu 
u_{\nu\nu}({\bf r})\right]^2+ \mu\sum_{\nu\nu'}\left[ u_{\nu\nu'}({\bf r})\right]^2\right\},
\eeq
where $\mu$ and $\lambda$ are  the Lam\'e parameters of the material, which 
can be inferred from experiments and/or from first principle calculations. In this paper 
we use the parameter values proposed in Ref.~\onlinecite{Atalaya08}, namely, 
$\mu = 103.89$ J/m$^2$ and $\lambda = 15.55$ J/m$^2$. 

This construction adds important elements to the purely geometric theory developed 
in Sec.~\ref{sec:out-plane}. Here, one needs the material parameters to establish a link
between the in-plane displacements $u_\nu$ and the topography $h$. We stress that
this procedure is still a geometric approach, since it incorporates $u_\nu$ in the metric.
This is different from the standard tight-binding theory, where the metric is entirely absent.

\begin{figure}[h]
\begin{center}
\centering \includegraphics[width=0.7\columnwidth]{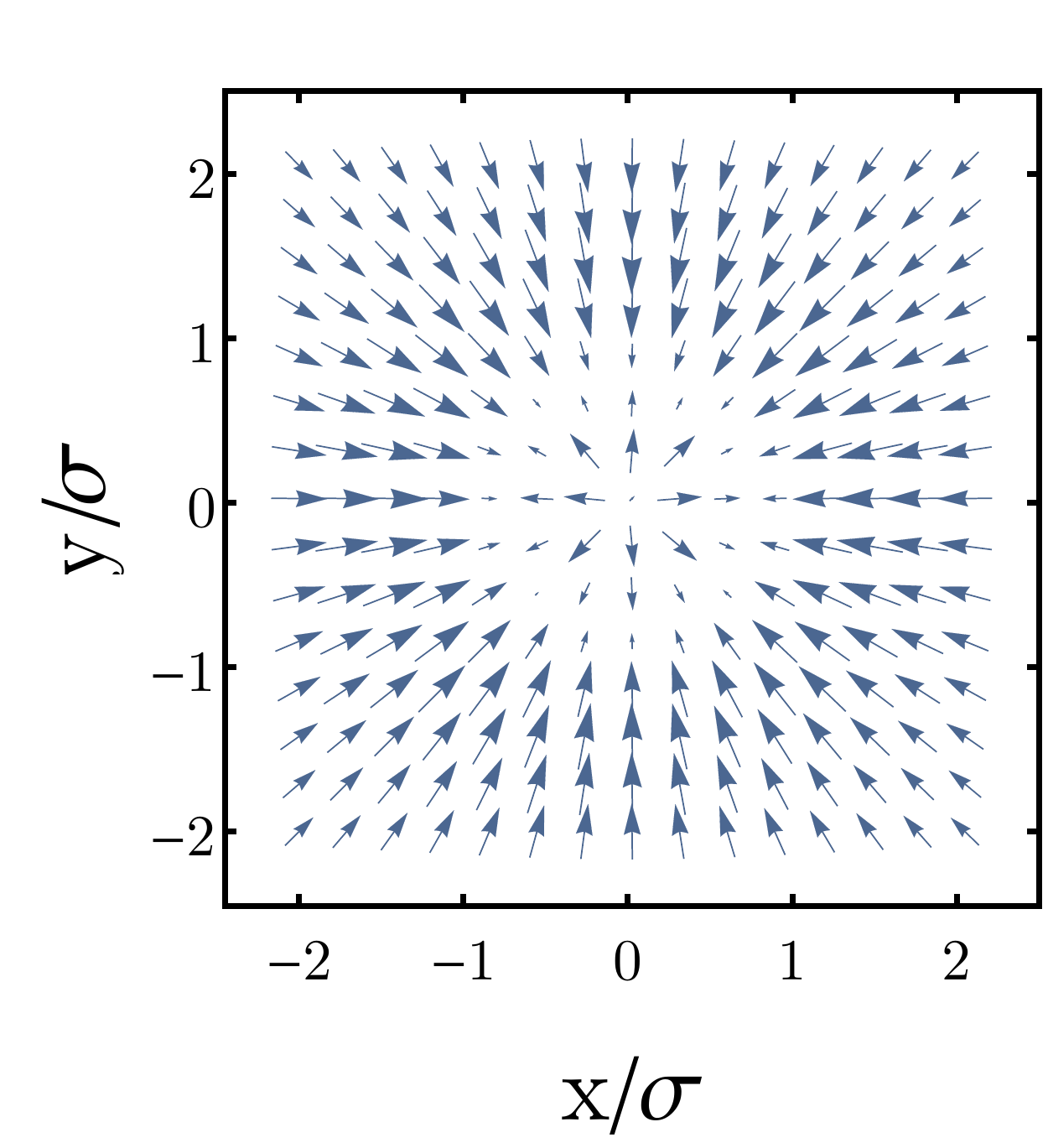} 
\caption{Vector field representing the in-plane displacement ${\bf u}(x, y)$ for a 
Gaussian bump with $h_0=1$ nm and $\sigma=5$ nm. The arrows represent the 
magnitude and direction of ${\bf u}$ (in arbitrary units). 
\label{fig:Dislocations}}
\end{center}
\end{figure}

Let us now calculate the in-plane displacement ${\bf u}(x,y)$ for the case of Gaussian 
deformation, namely, $h({\bf r})=h_0 \exp(-r^2/\sigma^2)$. This simple geometry allows 
for an analytical solution. 
The minimization of ${\cal H}_{\rm elastic}$  renders a set of differential equations 
that are solved in the momentum space \cite{Guinea08}. The result is
\beq
\bold u({\bf k})=-ih_0^2\,e^{-\frac{1}{8}k^2\sigma^2}
\frac{k^2\sigma^2(\lambda +2\mu)-8(\lambda +\mu)}{32k^2(\lambda +2\mu)}{\bf k},
\eeq
which, in configuration space, reads
\bea
\bold u({\bf{r}})&=&\sqrt{\frac{\pi}{2}}h_0^2\,\bold r\,\, e^{-2r^2/\sigma^2}\nn\\
&&\times \frac{-2r^2(\lambda +2\mu)+\sigma^2(\lambda +\mu)
\left(e^{2r^2/\sigma^2}-1\right)}{2r^2\sigma^2(\lambda +2\mu)}.\nn\\
\eea
The vector field representing the in-plane displacements ${\bf u}(x,y)$ is shown in 
Fig.~\ref{fig:Dislocations}.

The in-plane displacement ${\bf u}$ changes the strain tensor $u_{\nu\nu'}$, the metric, 
and consequently the local pseudo-magnetic field ${\cal B}_{\rm relax}$. 
Unfortunately, even for a topography as simple as that of a Gaussian bump, the analytical
expression for  ${\cal B}_{\rm relax}$ becomes rather lengthy and is not particularly 
insightful. Hence, we evaluate ${\cal B}_{\rm relax}$ numerically, following the steps 
described in Sec.~\ref{sec:in-and-out}.

\begin{figure}[h]
\begin{center}
\includegraphics[width=0.75\columnwidth]{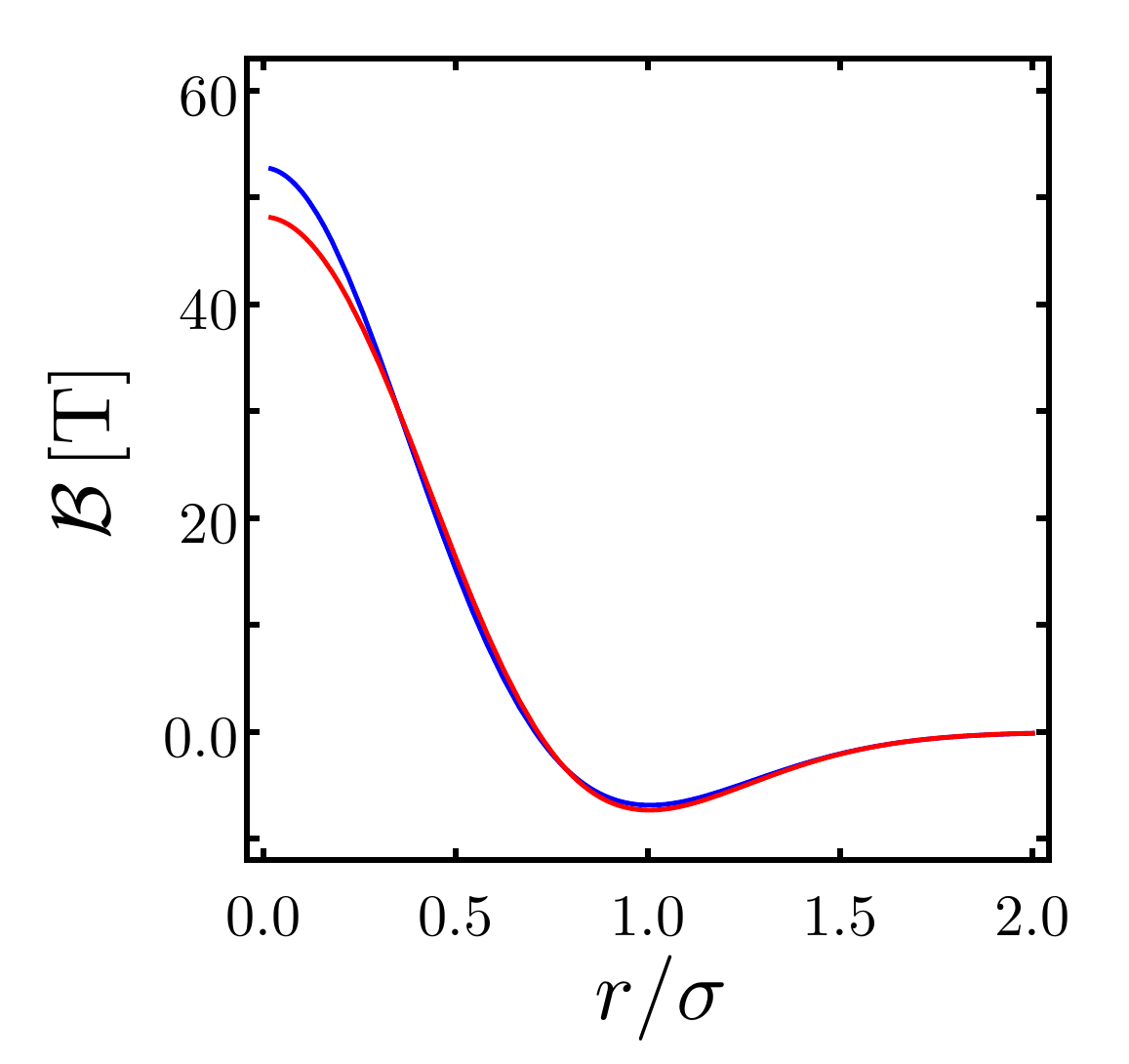} 
\caption{(Color online) Pseudo-magnetic field ${\cal B}$ as a function of $r$ 
corresponding to a Gaussian bump deformation with $h_0=1$ nm and 
$\sigma=5$ nm. The blue line represents the case without in-plane relaxation,
while the red curve accounts for both out-of-plane and in-plane deformations.  
\label{fig:Bvsr}}
\end{center}
\end{figure}

\begin{figure}[h]
\begin{center}
\includegraphics[width=0.65\columnwidth]{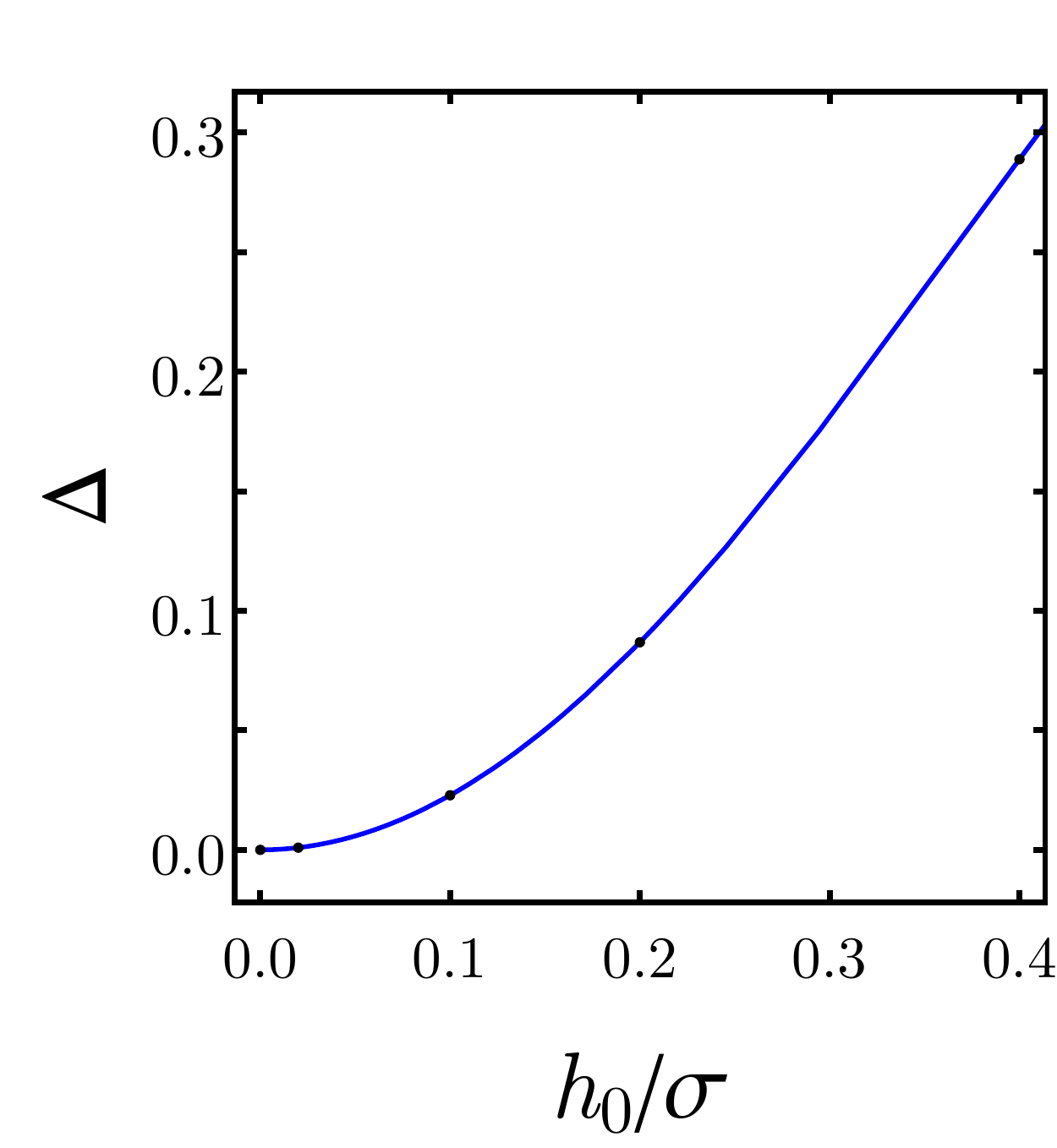} 
\caption{Ratio $\Delta$ between the in-plane contribution ${\cal B}-{\cal B}_{\rm relax}$ 
and the total pseudo-magnetic field ${\cal B}_{\rm relax}$ at ${\bf r} = 0$, defined by 
Eq.~\eqref{eq:def_zeta}, as a function of $h_0/\sigma$.}
\label{fig:DB/B}
\end{center}
\end{figure}

The physical picture that emerges is that the in-plane relaxation reduces the mechanical 
stress, diminishing the magnitude of the pseudo-magnetic field, as shown in Fig.~\ref{fig:Bvsr}.
We find that the in-plane relaxation corrections to ${\cal B}$ are small for ripples usually 
found in graphene deposited on standard substrates, where $h_0\ll \sigma$ 
\cite{Ishigami07,Geringer09}.
They become significative in situations where $h_0$ is comparable with $\sigma$, 
such as in the case of nanobubbles \cite{Levy2010}.
In Fig.~\ref{fig:DB/B} we plot a measure of the in-plane deformation contribution to the 
pseudo-magnetic field, namely, 
\beq
\label{eq:def_zeta}
\Delta=\left. \frac{{\cal B}-{\cal B}_{\rm relax}}{{\cal B}_{\rm relax}}\right|_{{\bf r}=0},
\eeq
as a function of $h_0/\sigma$.     

%
\section{Gauge invariance and the local plane approximation}
\label{sec:critique}

In the previous sections we have studied the influence of a modified metric on 
the Dirac Hamiltonian. So far, despite some elaborate efforts \cite{deJuan12,Yang2014}, 
a unified picture conciliating this quantum field theory approach with the 
standard band structure theory has not been established.
In this section we discuss some fundamental problems in constructing a 
bridge between these two approaches. From the practical point of view, 
as stressed in Ref.~\onlinecite{deJuan12}, while the geometric theory 
predicts a renormalization of the Fermi velocity, the elasticity approach 
does not. This issue was addressed in Ref.~\onlinecite{deJuan12}, where 
the authors analyze the local-flat Dirac equation starting from the 
global-curved one, namely
\beq
i\gamma^\mu(\partial_\mu+\Omega_\mu)\psi(\chi)=0,
\eeq
where $\gamma^\mu$ are the curved Dirac matrices 
$\gamma^\mu=\gamma^ae^\mu_{\,\,\,a}$. One can define 
the derivative with respect to the local-flat tangent space coordinates as
\beq 
\partial_a=\frac{\partial}{\partial\xi^a_{\cal X}}
e^{\mu}_{\,\,\,a}\partial_\mu .
\eeq 
%
The spin connection is given by $\Omega_\mu=\omega_{\mu12}\Sigma^{12}$ 
with $\Sigma^{12}=[\gamma^1,\gamma^2]/4$. 
By using the representation of the Dirac matrices for graphene
\beq\!
\gamma^0=\left(
\begin{array}{cc}
 \sigma_z & 0\\
 0 & \sigma_z
\end{array}
\right);
\hspace{0,09cm}
\gamma^1=i\left(
\begin{array}{cc}
 \sigma_x & 0\\
 0 & \sigma_x
\end{array}
\right);
\hspace{0,09cm}
\gamma^2=i\left(
\begin{array}{cc}
 \sigma_y & 0\\
 0 & -\sigma_y
\end{array}
\right)\!\!,
\eeq
where $\sigma_i$ are the  usual Pauli matrices, one obtains 
$\gamma^a\Sigma^{12}=-\frac{1}{2}\varepsilon^{ab}\gamma^{b}$. 
Thus, the Dirac Hamiltonian, in the local-flat tangent space, reads
\beq
i\gamma^a(\partial_a+A_a)\psi(x)=0,
\label{localdirac}
\eeq
where $A_a$ has been associated \cite{deJuan12} with a gauge potential defined by
\beq
A_a=\varepsilon_{ab}\,e^\mu_{\,\, b}\,\omega_{\mu12},
\label{Aflatlocal}
\eeq 
which originates the pseudo-magnetic field. 

Despite being very appealing, this construction of a Dirac equation with a gauge 
potential in flat-tangent space, is problematic. First, the gauge potential $A_a$ 
does not lead to a gauge invariant $B$-field..This statement is justified as follows: 
Let us assume we rotate the local-flat tangent space coordinates $\{\xi^a_{\cal X}\}$, 
defined at every point of the surface $\chi(x,y)$, by an angle $\delta\theta(x,y)$. 
The curved gauge 
potential ${\cal A}_\mu$  of Eq. (\ref{16}) and the local-flat potential $A_a$ transform 
as
\begin{align}
 \delta{\cal A}_\mu=&\partial_\mu(\delta\theta)    
 \quad \mbox{and} \quad
 \delta A_a=\varepsilon_{ab}\,e^\mu_{\,\, b}\partial_\mu(\delta\theta).
\end{align}
One can see that these local rotations constitute a gauge transformation for 
the curved potential ${\cal A}_\mu$ and therefore the corresponding 
pseudo-magnetic field is gauge invariant. Gauge invariance is not preserved 
for $A_a$,  since in the local-flat tangent space the pseudo-magnetic field given by 
$B=\varepsilon^{ab}\partial_aA_b$ depends explicitly on $\delta\theta$.

Secondly, the local-tangent space derivative $\partial_a$, Eq.~\eqref{localdirac}, 
is not a standard derivative,  since the commutator $[\partial_a,\partial_b]$ does 
not vanish, in general. We can show that
\beq
[\partial_a,\partial_b]=t_{ab}^{\,\,\,\,c}\partial_c,
\eeq
being the \textit{non-holonomicity coefficients}, $t_{ab}^{\,\,\,\,c}$, defined by 
\beq
 t_{ab}^{\,\,\,\,c}=\left(e^\mu_{\,\,\,a}\partial_\mu e^\nu_{\,\,\,b}-e^\mu_{\,\,\,b}\partial_\mu e^\nu_{\,\,\,a}\right)e_\nu^{\,\,\,c}.
\eeq 
This is different in the global-flat case, where $[\partial_\alpha,\partial_\beta]=0$, 
as discussed after Eq.~\eqref{eq:Dirac_flat}.
Therefore, the local-flat Dirac equation Eq. (\ref{localdirac}) can not be mapped into 
the usual Dirac equation in flat space. 

This problem remained unnoticed so far. Specifically Ref.~\onlinecite{Chaves14} 
uses similar vierbeins as ours, but with the condition $e_x^{\,\,\,2}=e_y^{\,\,\,1}$ 
to simplify the calculations. This choice implicitly fixes the gauge degree of freedom 
to be
\beq
\theta(x,y)=\arccos{\left(\frac{h_y^2+h_x^2\sqrt{1+h_ x^2+h_y^2}}{(h_x^2+h_y^2)
\sqrt{1+h_x^2}}\right)}.
\eeq
Due to this unphysical gauge dependence, the obtained pseudo-magnetic 
field is not unique.

\section{Conclusions}
\label{sec:conclusions}

In this paper we generalized the quantum field theory approach used to 
describe the low-energy electronic dynamics in rippled graphene. By 
considering a general two-dimensional curved graphene surface, we have 
properly identified the gauge transformations in that curved space. 
This lead us to define the effective gauge potential induced by the curvatures 
and to find an explicitly gauge invariant pseudo-magnetic field. We have found 
an equivalence between the pseudo-magnetic field induced by the ripples 
and the intrinsic scalar curvature of the general curved graphene surface. This 
remarkable relation, namely, ${\cal B}={\cal R}$, has been overlooked so far 
and constitutes the main contribution of this paper. 

We have extended these results to the realistic case where both 
out-of-plane and in-plane deformations are considered. 
We find that  this generalization preserves the equivalence between the 
pseudo-magnetic field and the intrinsic scalar curvature. 
As an application, we have analyzed the specific case of a Gaussian bump 
deformation.
We used the elasticity theory to write the deformation energy in terms of the 
strain tensor and few elastic material parameters. Following a minimization 
procedure \cite{Guinea08}, we found an analytical relation between the 
displacement field the system topography for a simple geometry.  
We analyzed the magnitude of the in-plane contributions to the pseudo-magnetic 
field. We found that since the pseudo-magnetic field is exactly 
the surface scalar curvature, it will be modified by in-plane deformations, this is because
these strains are not homogeneous nor uniform in space. 

We have also discussed the local-flat tangent space approximation used in the literature 
to identify the effective gauge field induced in deformed graphene. We conclude 
that the identification of a Dirac equation in that position-dependent tangent space is 
problematic, since it can lead to unphysical gauge dependent results.

\acknowledgments 

We thank Tobias Micklitz, Nami Svaiter, Luis Durand, and 
Marco Moriconi for useful discussions. This work has been 
supported by the Brazilian funding agencies CAPES, CNPq, 
and FAPERJ. 


\begin{thebibliography}{44}%
\makeatletter
\providecommand \@ifxundefined [1]{%
 \@ifx{#1\undefined}
}%
\providecommand \@ifnum [1]{%
 \ifnum #1\expandafter \@firstoftwo
 \else \expandafter \@secondoftwo
 \fi
}%
\providecommand \@ifx [1]{%
 \ifx #1\expandafter \@firstoftwo
 \else \expandafter \@secondoftwo
 \fi
}%
\providecommand \natexlab [1]{#1}%
\providecommand \enquote  [1]{``#1''}%
\providecommand \bibnamefont  [1]{#1}%
\providecommand \bibfnamefont [1]{#1}%
\providecommand \citenamefont [1]{#1}%
\providecommand \href@noop [0]{\@secondoftwo}%
\providecommand \href [0]{\begingroup \@sanitize@url \@href}%
\providecommand \@href[1]{\@@startlink{#1}\@@href}%
\providecommand \@@href[1]{\endgroup#1\@@endlink}%
\providecommand \@sanitize@url [0]{\catcode `\\12\catcode `\$12\catcode
  `\&12\catcode `\#12\catcode `\^12\catcode `\_12\catcode `\%12\relax}%
\providecommand \@@startlink[1]{}%
\providecommand \@@endlink[0]{}%
\providecommand \url  [0]{\begingroup\@sanitize@url \@url }%
\providecommand \@url [1]{\endgroup\@href {#1}{\urlprefix }}%
\providecommand \urlprefix  [0]{URL }%
\providecommand \Eprint [0]{\href }%
\providecommand \doibase [0]{http://dx.doi.org/}%
\providecommand \selectlanguage [0]{\@gobble}%
\providecommand \bibinfo  [0]{\@secondoftwo}%
\providecommand \bibfield  [0]{\@secondoftwo}%
\providecommand \translation [1]{[#1]}%
\providecommand \BibitemOpen [0]{}%
\providecommand \bibitemStop [0]{}%
\providecommand \bibitemNoStop [0]{.\EOS\space}%
\providecommand \EOS [0]{\spacefactor3000\relax}%
\providecommand \BibitemShut  [1]{\csname bibitem#1\endcsname}%
\let\auto@bib@innerbib\@empty
\bibitem [{\citenamefont {Bunch}\ \emph {et~al.}(2007)\citenamefont {Bunch},
  \citenamefont {van~der Zande}, \citenamefont {Verbridge}, \citenamefont
  {Frank}, \citenamefont {Tanenbaum}, \citenamefont {Parpia}, \citenamefont
  {Craighead},\ and\ \citenamefont {McEuen}}]{Bunch2007}%
  \BibitemOpen
  \bibfield  {author} {\bibinfo {author} {\bibfnamefont {J.~S.}\ \bibnamefont
  {Bunch}}, \bibinfo {author} {\bibfnamefont {A.~M.}\ \bibnamefont {van~der
  Zande}}, \bibinfo {author} {\bibfnamefont {S.~S.}\ \bibnamefont {Verbridge}},
  \bibinfo {author} {\bibfnamefont {I.~W.}\ \bibnamefont {Frank}}, \bibinfo
  {author} {\bibfnamefont {D.~M.}\ \bibnamefont {Tanenbaum}}, \bibinfo {author}
  {\bibfnamefont {J.~M.}\ \bibnamefont {Parpia}}, \bibinfo {author}
  {\bibfnamefont {H.~G.}\ \bibnamefont {Craighead}}, \ and\ \bibinfo {author}
  {\bibfnamefont {P.~L.}\ \bibnamefont {McEuen}},\ }\href {\doibase
  10.1126/science.1136836} {\bibfield  {journal} {\bibinfo  {journal}
  {Science}\ }\textbf {\bibinfo {volume} {315}},\ \bibinfo {pages} {490}
  (\bibinfo {year} {2007})}\BibitemShut {NoStop}%
\bibitem [{\citenamefont {Garcia-Sanchez}\ \emph {et~al.}(2008)\citenamefont
  {Garcia-Sanchez}, \citenamefont {van~der Zande}, \citenamefont {Paulo},
  \citenamefont {Lassagne}, \citenamefont {McEuen},\ and\ \citenamefont
  {Bachtold}}]{Garcia2008}%
  \BibitemOpen
  \bibfield  {author} {\bibinfo {author} {\bibfnamefont {D.}~\bibnamefont
  {Garcia-Sanchez}}, \bibinfo {author} {\bibfnamefont {A.~M.}\ \bibnamefont
  {van~der Zande}}, \bibinfo {author} {\bibfnamefont {A.~S.}\ \bibnamefont
  {Paulo}}, \bibinfo {author} {\bibfnamefont {B.}~\bibnamefont {Lassagne}},
  \bibinfo {author} {\bibfnamefont {P.~L.}\ \bibnamefont {McEuen}}, \ and\
  \bibinfo {author} {\bibfnamefont {A.}~\bibnamefont {Bachtold}},\ }\href
  {\doibase 10.1021/nl080201h} {\bibfield  {journal} {\bibinfo  {journal} {Nano
  Lett.}\ }\textbf {\bibinfo {volume} {8}},\ \bibinfo {pages} {1399} (\bibinfo
  {year} {2008})}\BibitemShut {NoStop}%
\bibitem [{\citenamefont {Chen}\ \emph {et~al.}(2009)\citenamefont {Chen},
  \citenamefont {Rosenblatt}, \citenamefont {Bolotin}, \citenamefont {Kalb},
  \citenamefont {Kim}, \citenamefont {Kymissis}, \citenamefont {Stormer},
  \citenamefont {Heinz},\ and\ \citenamefont {Hone}}]{Chen2009}%
  \BibitemOpen
  \bibfield  {author} {\bibinfo {author} {\bibfnamefont {C.}~\bibnamefont
  {Chen}}, \bibinfo {author} {\bibfnamefont {S.}~\bibnamefont {Rosenblatt}},
  \bibinfo {author} {\bibfnamefont {K.~I.}\ \bibnamefont {Bolotin}}, \bibinfo
  {author} {\bibfnamefont {W.}~\bibnamefont {Kalb}}, \bibinfo {author}
  {\bibfnamefont {P.}~\bibnamefont {Kim}}, \bibinfo {author} {\bibfnamefont
  {I.}~\bibnamefont {Kymissis}}, \bibinfo {author} {\bibfnamefont {H.~L.}\
  \bibnamefont {Stormer}}, \bibinfo {author} {\bibfnamefont {T.~F.}\
  \bibnamefont {Heinz}}, \ and\ \bibinfo {author} {\bibfnamefont
  {J.}~\bibnamefont {Hone}},\ }\href {http://dx.doi.org/10.1038/nnano.2009.267
  http://www.nature.com/nnano/journal/v4/n12/suppinfo/nnano.2009.267\_S1.html}
  {\bibfield  {journal} {\bibinfo  {journal} {Nature Nanotech.}\ }\textbf
  {\bibinfo {volume} {4}},\ \bibinfo {pages} {861} (\bibinfo {year}
  {2009})}\BibitemShut {NoStop}%
\bibitem [{\citenamefont {Eichler}\ \emph {et~al.}(2011)\citenamefont
  {Eichler}, \citenamefont {Moser}, \citenamefont {Chaste}, \citenamefont
  {Zdrojek}, \citenamefont {Wilson-Rae},\ and\ \citenamefont
  {Bachtold}}]{Eichler2011}%
  \BibitemOpen
  \bibfield  {author} {\bibinfo {author} {\bibfnamefont {A.}~\bibnamefont
  {Eichler}}, \bibinfo {author} {\bibfnamefont {J.}~\bibnamefont {Moser}},
  \bibinfo {author} {\bibfnamefont {J.}~\bibnamefont {Chaste}}, \bibinfo
  {author} {\bibfnamefont {M.}~\bibnamefont {Zdrojek}}, \bibinfo {author}
  {\bibfnamefont {I.}~\bibnamefont {Wilson-Rae}}, \ and\ \bibinfo {author}
  {\bibfnamefont {A.}~\bibnamefont {Bachtold}},\ }\href
  {http://dx.doi.org/10.1038/nnano.2011.71
  http://www.nature.com/nnano/journal/v6/n6/abs/nnano.2011.71.html\#supplementary-information}
  {\bibfield  {journal} {\bibinfo  {journal} {Nature Nanotech.}\ }\textbf
  {\bibinfo {volume} {6}},\ \bibinfo {pages} {339} (\bibinfo {year}
  {2011})}\BibitemShut {NoStop}%
\bibitem [{\citenamefont {Jones}\ and\ \citenamefont
  {Pereira}(2014)}]{Jones2014}%
  \BibitemOpen
  \bibfield  {author} {\bibinfo {author} {\bibfnamefont {G.~W.}\ \bibnamefont
  {Jones}}\ and\ \bibinfo {author} {\bibfnamefont {V.~M.}\ \bibnamefont
  {Pereira}},\ }\href {\doibase 10.1088/1367-2630/16/9/093044} {\bibfield
  {journal} {\bibinfo  {journal} {New J. Phys.}\ }\textbf {\bibinfo {volume}
  {16}},\ \bibinfo {pages} {093044} (\bibinfo {year} {2014})}\BibitemShut
  {NoStop}%
\bibitem [{\citenamefont {Guinea}\ \emph {et~al.}(2009)\citenamefont {Guinea},
  \citenamefont {Katsnelson},\ and\ \citenamefont {Geim}}]{Guinea2009}%
  \BibitemOpen
  \bibfield  {author} {\bibinfo {author} {\bibfnamefont {F.}~\bibnamefont
  {Guinea}}, \bibinfo {author} {\bibfnamefont {M.~I.}\ \bibnamefont
  {Katsnelson}}, \ and\ \bibinfo {author} {\bibfnamefont {A.~K.}\ \bibnamefont
  {Geim}},\ }\href {\doibase 10.1038/nphys1420} {\bibfield  {journal} {\bibinfo
   {journal} {Nature Phys.}\ }\textbf {\bibinfo {volume} {6}},\ \bibinfo
  {pages} {30} (\bibinfo {year} {2009})}\BibitemShut {NoStop}%
\bibitem [{\citenamefont {Pereira}\ and\ \citenamefont {{Castro
  Neto}}(2009)}]{Pereira09b}%
  \BibitemOpen
  \bibfield  {author} {\bibinfo {author} {\bibfnamefont {V.~M.}\ \bibnamefont
  {Pereira}}\ and\ \bibinfo {author} {\bibfnamefont {A.~H.}\ \bibnamefont
  {{Castro Neto}}},\ }\href {\doibase 10.1103/PhysRevLett.103.046801}
  {\bibfield  {journal} {\bibinfo  {journal} {Phys. Rev. Lett.}\ }\textbf
  {\bibinfo {volume} {103}},\ \bibinfo {pages} {046801} (\bibinfo {year}
  {2009})}\BibitemShut {NoStop}%
\bibitem [{\citenamefont {Low}\ \emph {et~al.}(2011)\citenamefont {Low},
  \citenamefont {Guinea},\ and\ \citenamefont {Katsnelson}}]{Low2011}%
  \BibitemOpen
  \bibfield  {author} {\bibinfo {author} {\bibfnamefont {T.}~\bibnamefont
  {Low}}, \bibinfo {author} {\bibfnamefont {F.}~\bibnamefont {Guinea}}, \ and\
  \bibinfo {author} {\bibfnamefont {M.~I.}\ \bibnamefont {Katsnelson}},\ }\href
  {\doibase 10.1103/PhysRevB.83.195436} {\bibfield  {journal} {\bibinfo
  {journal} {Phys. Rev. B}\ }\textbf {\bibinfo {volume} {83}},\ \bibinfo
  {pages} {195436} (\bibinfo {year} {2011})}\BibitemShut {NoStop}%
\bibitem [{\citenamefont {Prada}\ \emph {et~al.}(2009)\citenamefont {Prada},
  \citenamefont {San-Jose},\ and\ \citenamefont {Schomerus}}]{Prada2009}%
  \BibitemOpen
  \bibfield  {author} {\bibinfo {author} {\bibfnamefont {E.}~\bibnamefont
  {Prada}}, \bibinfo {author} {\bibfnamefont {P.}~\bibnamefont {San-Jose}}, \
  and\ \bibinfo {author} {\bibfnamefont {H.}~\bibnamefont {Schomerus}},\ }\href
  {\doibase 10.1103/PhysRevB.80.245414} {\bibfield  {journal} {\bibinfo
  {journal} {Phys. Rev. B}\ }\textbf {\bibinfo {volume} {80}},\ \bibinfo
  {pages} {245414} (\bibinfo {year} {2009})}\BibitemShut {NoStop}%
\bibitem [{\citenamefont {Low}\ \emph {et~al.}(2012)\citenamefont {Low},
  \citenamefont {Jiang}, \citenamefont {Katsnelson},\ and\ \citenamefont
  {Guinea}}]{Low12}%
  \BibitemOpen
  \bibfield  {author} {\bibinfo {author} {\bibfnamefont {T.}~\bibnamefont
  {Low}}, \bibinfo {author} {\bibfnamefont {Y.}~\bibnamefont {Jiang}}, \bibinfo
  {author} {\bibfnamefont {M.~I.}\ \bibnamefont {Katsnelson}}, \ and\ \bibinfo
  {author} {\bibfnamefont {F.}~\bibnamefont {Guinea}},\ }\href@noop {}
  {\bibfield  {journal} {\bibinfo  {journal} {Nano Lett.}\ }\textbf {\bibinfo
  {volume} {12}},\ \bibinfo {pages} {850} (\bibinfo {year} {2012})}\BibitemShut
  {NoStop}%
\bibitem [{\citenamefont {Jiang}\ \emph {et~al.}(2013)\citenamefont {Jiang},
  \citenamefont {Low}, \citenamefont {Chang}, \citenamefont {Katsnelson},\ and\
  \citenamefont {Guinea}}]{Jiang2013}%
  \BibitemOpen
  \bibfield  {author} {\bibinfo {author} {\bibfnamefont {Y.}~\bibnamefont
  {Jiang}}, \bibinfo {author} {\bibfnamefont {T.}~\bibnamefont {Low}}, \bibinfo
  {author} {\bibfnamefont {K.}~\bibnamefont {Chang}}, \bibinfo {author}
  {\bibfnamefont {M.~I.}\ \bibnamefont {Katsnelson}}, \ and\ \bibinfo {author}
  {\bibfnamefont {F.}~\bibnamefont {Guinea}},\ }\href@noop {} {\bibfield
  {journal} {\bibinfo  {journal} {Phys. Rev. Lett.}\ }\textbf {\bibinfo
  {volume} {110}},\ \bibinfo {pages} {046601} (\bibinfo {year}
  {2013})}\BibitemShut {NoStop}%
\bibitem [{\citenamefont {Zhu}\ \emph {et~al.}(2014)\citenamefont {Zhu},
  \citenamefont {Huang}, \citenamefont {Klimov}, \citenamefont {Newell},
  \citenamefont {Zhitenev}, \citenamefont {Stroscio}, \citenamefont {Solares},\
  and\ \citenamefont {Li}}]{Zhu2014}%
  \BibitemOpen
  \bibfield  {author} {\bibinfo {author} {\bibfnamefont {S.}~\bibnamefont
  {Zhu}}, \bibinfo {author} {\bibfnamefont {Y.}~\bibnamefont {Huang}}, \bibinfo
  {author} {\bibfnamefont {N.~N.}\ \bibnamefont {Klimov}}, \bibinfo {author}
  {\bibfnamefont {D.~B.}\ \bibnamefont {Newell}}, \bibinfo {author}
  {\bibfnamefont {N.~B.}\ \bibnamefont {Zhitenev}}, \bibinfo {author}
  {\bibfnamefont {J.~A.}\ \bibnamefont {Stroscio}}, \bibinfo {author}
  {\bibfnamefont {S.~D.}\ \bibnamefont {Solares}}, \ and\ \bibinfo {author}
  {\bibfnamefont {T.}~\bibnamefont {Li}},\ }\href {\doibase
  10.1103/PhysRevB.90.075426} {\bibfield  {journal} {\bibinfo  {journal} {Phys.
  Rev. B}\ }\textbf {\bibinfo {volume} {90}},\ \bibinfo {pages} {075426}
  (\bibinfo {year} {2014})}\BibitemShut {NoStop}%
\bibitem [{\citenamefont {Carrillo-Bastos}\ \emph {et~al.}(2014)\citenamefont
  {Carrillo-Bastos}, \citenamefont {Faria}, \citenamefont {Latge},
  \citenamefont {Mireles},\ and\ \citenamefont {Sandler}}]{Bastos2014}%
  \BibitemOpen
  \bibfield  {author} {\bibinfo {author} {\bibfnamefont {R.}~\bibnamefont
  {Carrillo-Bastos}}, \bibinfo {author} {\bibfnamefont {D.}~\bibnamefont
  {Faria}}, \bibinfo {author} {\bibfnamefont {A.}~\bibnamefont {Latge}},
  \bibinfo {author} {\bibfnamefont {F.}~\bibnamefont {Mireles}}, \ and\
  \bibinfo {author} {\bibfnamefont {N.}~\bibnamefont {Sandler}},\ }\href
  {\doibase 10.1103/PhysRevB.90.041411} {\bibfield  {journal} {\bibinfo
  {journal} {Phys. Rev. B}\ }\textbf {\bibinfo {volume} {90}},\ \bibinfo
  {pages} {041411(R)} (\bibinfo {year} {2014})}\BibitemShut {NoStop}%
\bibitem [{\citenamefont {Suzuura}\ and\ \citenamefont
  {Ando}(2002)}]{Suzuura2002}%
  \BibitemOpen
  \bibfield  {author} {\bibinfo {author} {\bibfnamefont {H.}~\bibnamefont
  {Suzuura}}\ and\ \bibinfo {author} {\bibfnamefont {T.}~\bibnamefont {Ando}},\
  }\href {\doibase 10.1103/PhysRevB.65.235412} {\bibfield  {journal} {\bibinfo
  {journal} {Phys. Rev. B}\ }\textbf {\bibinfo {volume} {65}},\ \bibinfo
  {pages} {235412} (\bibinfo {year} {2002})}\BibitemShut {NoStop}%
\bibitem [{\citenamefont {Ma\~{n}es}(2007)}]{Manes2007}%
  \BibitemOpen
  \bibfield  {author} {\bibinfo {author} {\bibfnamefont {J.~L.}\ \bibnamefont
  {Ma\~{n}es}},\ }\href {\doibase 10.1103/PhysRevB.76.045430} {\bibfield
  {journal} {\bibinfo  {journal} {Phys. Rev. B}\ }\textbf {\bibinfo {volume}
  {76}},\ \bibinfo {pages} {045430} (\bibinfo {year} {2007})}\BibitemShut
  {NoStop}%
\bibitem [{\citenamefont {Guinea}\ \emph {et~al.}(2008)\citenamefont {Guinea},
  \citenamefont {Horovitz},\ and\ \citenamefont {{Le Doussal}}}]{Guinea08}%
  \BibitemOpen
  \bibfield  {author} {\bibinfo {author} {\bibfnamefont {F.}~\bibnamefont
  {Guinea}}, \bibinfo {author} {\bibfnamefont {B.}~\bibnamefont {Horovitz}}, \
  and\ \bibinfo {author} {\bibfnamefont {P.}~\bibnamefont {{Le Doussal}}},\
  }\href {\doibase 10.1103/PhysRevB.77.205421} {\bibfield  {journal} {\bibinfo
  {journal} {Phys. Rev. B}\ }\textbf {\bibinfo {volume} {77}},\ \bibinfo
  {pages} {205421} (\bibinfo {year} {2008})}\BibitemShut {NoStop}%
\bibitem [{\citenamefont {Ma\~{n}es}\ \emph {et~al.}(2013)\citenamefont
  {Ma\~{n}es}, \citenamefont {de~Juan}, \citenamefont {Sturla},\ and\
  \citenamefont {Vozmediano}}]{Manes2013}%
  \BibitemOpen
  \bibfield  {author} {\bibinfo {author} {\bibfnamefont {J.~L.}\ \bibnamefont
  {Ma\~{n}es}}, \bibinfo {author} {\bibfnamefont {F.}~\bibnamefont {de~Juan}},
  \bibinfo {author} {\bibfnamefont {M.}~\bibnamefont {Sturla}}, \ and\ \bibinfo
  {author} {\bibfnamefont {M.~A.~H.}\ \bibnamefont {Vozmediano}},\ }\href
  {\doibase 10.1103/PhysRevB.88.155405} {\bibfield  {journal} {\bibinfo
  {journal} {Phys. Rev. B}\ }\textbf {\bibinfo {volume} {88}},\ \bibinfo
  {pages} {155405} (\bibinfo {year} {2013})}\BibitemShut {NoStop}%
\bibitem [{\citenamefont {de~Juan}\ \emph {et~al.}(2013)\citenamefont
  {de~Juan}, \citenamefont {Ma\~{n}es},\ and\ \citenamefont
  {Vozmediano}}]{deJuan13}%
  \BibitemOpen
  \bibfield  {author} {\bibinfo {author} {\bibfnamefont {F.}~\bibnamefont
  {de~Juan}}, \bibinfo {author} {\bibfnamefont {J.~L.}\ \bibnamefont
  {Ma\~{n}es}}, \ and\ \bibinfo {author} {\bibfnamefont {M.~A.~H.}\
  \bibnamefont {Vozmediano}},\ }\href {\doibase 10.1103/PhysRevB.87.165131}
  {\bibfield  {journal} {\bibinfo  {journal} {Phys. Rev. B}\ }\textbf {\bibinfo
  {volume} {87}},\ \bibinfo {pages} {165131} (\bibinfo {year}
  {2013})}\BibitemShut {NoStop}%
\bibitem [{\citenamefont {{Ramezani Masir}}\ \emph {et~al.}(2013)\citenamefont
  {{Ramezani Masir}}, \citenamefont {Moldovan},\ and\ \citenamefont
  {Peeters}}]{Masir2013}%
  \BibitemOpen
  \bibfield  {author} {\bibinfo {author} {\bibfnamefont {M.}~\bibnamefont
  {{Ramezani Masir}}}, \bibinfo {author} {\bibfnamefont {D.}~\bibnamefont
  {Moldovan}}, \ and\ \bibinfo {author} {\bibfnamefont {F.}~\bibnamefont
  {Peeters}},\ }\href {\doibase http://dx.doi.org/10.1016/j.ssc.2013.04.001}
  {\bibfield  {journal} {\bibinfo  {journal} {Solid State Commun.}\ }\textbf
  {\bibinfo {volume} {175-€"176}},\ \bibinfo {pages} {76 } (\bibinfo {year}
  {2013})}\BibitemShut {NoStop}%
\bibitem [{\citenamefont {Levy}\ \emph {et~al.}(2010)\citenamefont {Levy},
  \citenamefont {Burke}, \citenamefont {Meaker}, \citenamefont {Panlasigui},
  \citenamefont {Zettl}, \citenamefont {Guinea}, \citenamefont {{Castro
  Neto}},\ and\ \citenamefont {Crommie}}]{Levy2010}%
  \BibitemOpen
  \bibfield  {author} {\bibinfo {author} {\bibfnamefont {N.}~\bibnamefont
  {Levy}}, \bibinfo {author} {\bibfnamefont {S.~A.}\ \bibnamefont {Burke}},
  \bibinfo {author} {\bibfnamefont {K.~L.}\ \bibnamefont {Meaker}}, \bibinfo
  {author} {\bibfnamefont {M.}~\bibnamefont {Panlasigui}}, \bibinfo {author}
  {\bibfnamefont {A.}~\bibnamefont {Zettl}}, \bibinfo {author} {\bibfnamefont
  {F.}~\bibnamefont {Guinea}}, \bibinfo {author} {\bibfnamefont {A.~H.}\
  \bibnamefont {{Castro Neto}}}, \ and\ \bibinfo {author} {\bibfnamefont
  {M.~F.}\ \bibnamefont {Crommie}},\ }\href {\doibase 10.1126/science.1191700}
  {\bibfield  {journal} {\bibinfo  {journal} {Science}\ }\textbf {\bibinfo
  {volume} {329}},\ \bibinfo {pages} {544} (\bibinfo {year}
  {2010})}\BibitemShut {NoStop}%
\bibitem [{\citenamefont {Yeh}\ \emph {et~al.}(2011)\citenamefont {Yeh},
  \citenamefont {Teague}, \citenamefont {Yeom}, \citenamefont {Standley},
  \citenamefont {Wu}, \citenamefont {Boyd},\ and\ \citenamefont
  {Bockrath}}]{Yeh2011}%
  \BibitemOpen
  \bibfield  {author} {\bibinfo {author} {\bibfnamefont {N.-C.}\ \bibnamefont
  {Yeh}}, \bibinfo {author} {\bibfnamefont {M.-L.}\ \bibnamefont {Teague}},
  \bibinfo {author} {\bibfnamefont {S.}~\bibnamefont {Yeom}}, \bibinfo {author}
  {\bibfnamefont {B.~L.}\ \bibnamefont {Standley}}, \bibinfo {author}
  {\bibfnamefont {R.-P.}\ \bibnamefont {Wu}}, \bibinfo {author} {\bibfnamefont
  {D.~A.}\ \bibnamefont {Boyd}}, \ and\ \bibinfo {author} {\bibfnamefont
  {M.~W.}\ \bibnamefont {Bockrath}},\ }\href {\doibase
  http://dx.doi.org/10.1016/j.susc.2011.03.025} {\bibfield  {journal} {\bibinfo
   {journal} {Surface Science}\ }\textbf {\bibinfo {volume} {605}},\ \bibinfo
  {pages} {1649} (\bibinfo {year} {2011})}\BibitemShut {NoStop}%
\bibitem [{\citenamefont {Midtvedt}\ \emph {et~al.}(2015)\citenamefont
  {Midtvedt}, \citenamefont {Lewenkopf},\ and\ \citenamefont
  {Croy}}]{Midtvedt2015}%
  \BibitemOpen
  \bibfield  {author} {\bibinfo {author} {\bibfnamefont {D.}~\bibnamefont
  {Midtvedt}}, \bibinfo {author} {\bibfnamefont {C.~H.}\ \bibnamefont
  {Lewenkopf}}, \ and\ \bibinfo {author} {\bibfnamefont {A.}~\bibnamefont
  {Croy}},\ }\href@noop {} {\bibfield  {journal} {\bibinfo  {journal}
  {arXiv:1509.02365}\ } (\bibinfo {year} {2015})}\BibitemShut {NoStop}%
\bibitem [{\citenamefont {Lundeberg}\ and\ \citenamefont
  {Folk}(2010)}]{Lundeberg10}%
  \BibitemOpen
  \bibfield  {author} {\bibinfo {author} {\bibfnamefont {M.~B.}\ \bibnamefont
  {Lundeberg}}\ and\ \bibinfo {author} {\bibfnamefont {J.~A.}\ \bibnamefont
  {Folk}},\ }\href {\doibase 10.1103/PhysRevLett.105.146804} {\bibfield
  {journal} {\bibinfo  {journal} {Phys. Rev. Lett.}\ }\textbf {\bibinfo
  {volume} {105}},\ \bibinfo {pages} {146804} (\bibinfo {year}
  {2010})}\BibitemShut {NoStop}%
\bibitem [{\citenamefont {Burgos}\ \emph {et~al.}(2015)\citenamefont {Burgos},
  \citenamefont {Warnes}, \citenamefont {Lima},\ and\ \citenamefont
  {Lewenkopf}}]{Burgos2015}%
  \BibitemOpen
  \bibfield  {author} {\bibinfo {author} {\bibfnamefont {R.}~\bibnamefont
  {Burgos}}, \bibinfo {author} {\bibfnamefont {J.}~\bibnamefont {Warnes}},
  \bibinfo {author} {\bibfnamefont {L.~R.~F.}\ \bibnamefont {Lima}}, \ and\
  \bibinfo {author} {\bibfnamefont {C.}~\bibnamefont {Lewenkopf}},\ }\href
  {\doibase 10.1103/PhysRevB.91.115403} {\bibfield  {journal} {\bibinfo
  {journal} {Phys. Rev. B}\ }\textbf {\bibinfo {volume} {91}},\ \bibinfo
  {pages} {115403} (\bibinfo {year} {2015})}\BibitemShut {NoStop}%
\bibitem [{\citenamefont {Oliva-Leyva}\ and\ \citenamefont
  {Naumis}(2013)}]{OlivaLeyva2013}%
  \BibitemOpen
  \bibfield  {author} {\bibinfo {author} {\bibfnamefont {M.}~\bibnamefont
  {Oliva-Leyva}}\ and\ \bibinfo {author} {\bibfnamefont {G.~G.}\ \bibnamefont
  {Naumis}},\ }\href {\doibase 10.1103/PhysRevB.88.085430} {\bibfield
  {journal} {\bibinfo  {journal} {Phys. Rev. B}\ }\textbf {\bibinfo {volume}
  {88}},\ \bibinfo {pages} {085430} (\bibinfo {year} {2013})}\BibitemShut
  {NoStop}%
\bibitem [{\citenamefont {Weinberg}(1972)}]{Weinberg1972}%
  \BibitemOpen
  \bibfield  {author} {\bibinfo {author} {\bibfnamefont {S.}~\bibnamefont
  {Weinberg}},\ }\href@noop {} {\emph {\bibinfo {title} {{Gravitation and
  Cosmology: Principles and Applications to the General Theory of
  Relativity}}}}\ (\bibinfo  {publisher} {Wiley and Sons},\ \bibinfo {address}
  {New York},\ \bibinfo {year} {1972})\BibitemShut {NoStop}%
\bibitem [{\citenamefont {Birrell}\ and\ \citenamefont
  {Davis}(1982)}]{Birrell1982}%
  \BibitemOpen
  \bibfield  {author} {\bibinfo {author} {\bibfnamefont {N.~D.}\ \bibnamefont
  {Birrell}}\ and\ \bibinfo {author} {\bibfnamefont {P.~C.~W.}\ \bibnamefont
  {Davis}},\ }\href@noop {} {\emph {\bibinfo {title} {{Quantum Fields in Curved
  Space}}}}\ (\bibinfo  {publisher} {Cambridge University Press},\ \bibinfo
  {address} {Cambridge},\ \bibinfo {year} {1982})\BibitemShut {NoStop}%
\bibitem [{\citenamefont {Semenoff}(1984)}]{Semenoff84}%
  \BibitemOpen
  \bibfield  {author} {\bibinfo {author} {\bibfnamefont {G.~W.}\ \bibnamefont
  {Semenoff}},\ }\href {\doibase 10.1103/PhysRevLett.53.2449} {\bibfield
  {journal} {\bibinfo  {journal} {Phys. Rev. Lett.}\ }\textbf {\bibinfo
  {volume} {53}},\ \bibinfo {pages} {2449} (\bibinfo {year}
  {1984})}\BibitemShut {NoStop}%
\bibitem [{\citenamefont {Vozmediano}\ \emph {et~al.}(2010)\citenamefont
  {Vozmediano}, \citenamefont {Katsnelson},\ and\ \citenamefont
  {Guinea}}]{Vozmediano10}%
  \BibitemOpen
  \bibfield  {author} {\bibinfo {author} {\bibfnamefont {M.~A.~H.}\
  \bibnamefont {Vozmediano}}, \bibinfo {author} {\bibfnamefont {M.~I.}\
  \bibnamefont {Katsnelson}}, \ and\ \bibinfo {author} {\bibfnamefont
  {F.}~\bibnamefont {Guinea}},\ }\href {\doibase 10.1016/j.physrep.2010.07.003}
  {\bibfield  {journal} {\bibinfo  {journal} {Phys. Rep.}\ }\textbf {\bibinfo
  {volume} {496}},\ \bibinfo {pages} {109} (\bibinfo {year}
  {2010})}\BibitemShut {NoStop}%
\bibitem [{\citenamefont {de~Juan}\ \emph {et~al.}(2007)\citenamefont
  {de~Juan}, \citenamefont {Cortijo},\ and\ \citenamefont
  {Vozmediano}}]{deJuan07}%
  \BibitemOpen
  \bibfield  {author} {\bibinfo {author} {\bibfnamefont {F.}~\bibnamefont
  {de~Juan}}, \bibinfo {author} {\bibfnamefont {A.}~\bibnamefont {Cortijo}}, \
  and\ \bibinfo {author} {\bibfnamefont {M.~A.~H.}\ \bibnamefont
  {Vozmediano}},\ }\href {\doibase 10.1103/PhysRevB.76.165409} {\bibfield
  {journal} {\bibinfo  {journal} {Phys. Rev. B}\ }\textbf {\bibinfo {volume}
  {76}},\ \bibinfo {pages} {165409} (\bibinfo {year} {2007})}\BibitemShut
  {NoStop}%
\bibitem [{\citenamefont {Yan}\ \emph {et~al.}(2013)\citenamefont {Yan},
  \citenamefont {Chu}, \citenamefont {Yan}, \citenamefont {Liu}, \citenamefont
  {Meng}, \citenamefont {Yang}, \citenamefont {Fan}, \citenamefont {Wang},
  \citenamefont {Dou}, \citenamefont {Zhang}, \citenamefont {Liu},
  \citenamefont {Nie},\ and\ \citenamefont {He}}]{Yan2013}%
  \BibitemOpen
  \bibfield  {author} {\bibinfo {author} {\bibfnamefont {H.}~\bibnamefont
  {Yan}}, \bibinfo {author} {\bibfnamefont {Z.-D.}\ \bibnamefont {Chu}},
  \bibinfo {author} {\bibfnamefont {W.}~\bibnamefont {Yan}}, \bibinfo {author}
  {\bibfnamefont {M.}~\bibnamefont {Liu}}, \bibinfo {author} {\bibfnamefont
  {L.}~\bibnamefont {Meng}}, \bibinfo {author} {\bibfnamefont {M.}~\bibnamefont
  {Yang}}, \bibinfo {author} {\bibfnamefont {Y.}~\bibnamefont {Fan}}, \bibinfo
  {author} {\bibfnamefont {J.}~\bibnamefont {Wang}}, \bibinfo {author}
  {\bibfnamefont {R.-F.}\ \bibnamefont {Dou}}, \bibinfo {author} {\bibfnamefont
  {Y.}~\bibnamefont {Zhang}}, \bibinfo {author} {\bibfnamefont
  {Z.}~\bibnamefont {Liu}}, \bibinfo {author} {\bibfnamefont {J.-C.}\
  \bibnamefont {Nie}}, \ and\ \bibinfo {author} {\bibfnamefont
  {L.}~\bibnamefont {He}},\ }\href {\doibase 10.1103/PhysRevB.87.075405}
  {\bibfield  {journal} {\bibinfo  {journal} {Phys. Rev. B}\ }\textbf {\bibinfo
  {volume} {87}},\ \bibinfo {pages} {075405} (\bibinfo {year}
  {2013})}\BibitemShut {NoStop}%
\bibitem [{\citenamefont {Sanjuan}\ \emph {et~al.}(2014)\citenamefont
  {Sanjuan}, \citenamefont {Wang}, \citenamefont {Imani}, \citenamefont
  {Vanevic},\ and\ \citenamefont {Barraza-Lopez}}]{Sanjuan2014a}%
  \BibitemOpen
  \bibfield  {author} {\bibinfo {author} {\bibfnamefont {A.~A.}\
  \bibnamefont {Pacheco Sanjuan}}, \bibinfo {author} {\bibfnamefont {Z.}~\bibnamefont
  {Wang}}, \bibinfo {author} {\bibfnamefont {H.}\ \bibnamefont {Pour Imani}},
  \bibinfo {author} {\bibfnamefont {M.}~\bibnamefont {Vanevic}}, \ and\
  \bibinfo {author} {\bibfnamefont {S.}~\bibnamefont {Barraza-Lopez}},\ }\href
  {\doibase 10.1103/PhysRevB.89.121403} {\bibfield  {journal} {\bibinfo
  {journal} {Phys. Rev. B}\ }\textbf {\bibinfo {volume} {89}},\ \bibinfo
  {pages} {121403(R)} (\bibinfo {year} {2014})}\BibitemShut {NoStop}%
\bibitem [{\citenamefont {Khveshchenko}(2013)}]{Khveshchenko2013}%
  \BibitemOpen
  \bibfield  {author} {\bibinfo {author} {\bibfnamefont {D.~V.}\ \bibnamefont
  {Khveshchenko}},\ }\href {\doibase 10.1209/0295-5075/104/47002} {\bibfield
  {journal} {\bibinfo  {journal} {EPL (Europhysics Letters)}\ }\textbf
  {\bibinfo {volume} {104}},\ \bibinfo {pages} {47002} (\bibinfo {year}
  {2013})}\BibitemShut {NoStop}%
\bibitem [{\citenamefont {Volovik}\ and\ \citenamefont
  {Zubkov}(2014)}]{Volovik2014}%
  \BibitemOpen
  \bibfield  {author} {\bibinfo {author} {\bibfnamefont {G.~E.}\ \bibnamefont
  {Volovik}}\ and\ \bibinfo {author} {\bibfnamefont {M.~A.}\ \bibnamefont
  {Zubkov}},\ }\href {\doibase 10.1016/j.nuclphysb.2014.02.018} {\bibfield
  {journal} {\bibinfo  {journal} {Nucl. Phys. B}\ }\textbf {\bibinfo {volume}
  {881}},\ \bibinfo {pages} {514} (\bibinfo {year} {2014})}\BibitemShut
  {NoStop}%
\bibitem [{\citenamefont {Landau}\ and\ \citenamefont
  {Lifshitz}(1986)}]{Landau1986}%
  \BibitemOpen
  \bibfield  {author} {\bibinfo {author} {\bibfnamefont {L.~D.}\ \bibnamefont
  {Landau}}\ and\ \bibinfo {author} {\bibfnamefont {E.~M.}\ \bibnamefont
  {Lifshitz}},\ }\href@noop {} {\emph {\bibinfo {title} {{Theory of
  elasticity}}}},\ \bibinfo {edition} {3rd}\ ed.\ (\bibinfo  {publisher}
  {Butterworth-Heinemann},\ \bibinfo {address} {Oxford},\ \bibinfo {year}
  {1986})\BibitemShut {NoStop}%
\bibitem [{\citenamefont {de~Juan}\ \emph {et~al.}(2012)\citenamefont
  {de~Juan}, \citenamefont {Sturla},\ and\ \citenamefont
  {Vozmediano}}]{deJuan12}%
  \BibitemOpen
  \bibfield  {author} {\bibinfo {author} {\bibfnamefont {F.}~\bibnamefont
  {de~Juan}}, \bibinfo {author} {\bibfnamefont {M.}~\bibnamefont {Sturla}}, \
  and\ \bibinfo {author} {\bibfnamefont {M.~A.~H.}\ \bibnamefont
  {Vozmediano}},\ }\href {\doibase 10.1103/PhysRevLett.108.227205} {\bibfield
  {journal} {\bibinfo  {journal} {Phys. Rev. Lett.}\ }\textbf {\bibinfo
  {volume} {108}},\ \bibinfo {pages} {227205} (\bibinfo {year}
  {2012})}\BibitemShut {NoStop}%
\bibitem [{\citenamefont {Chaves}\ \emph {et~al.}(2014)\citenamefont {Chaves},
  \citenamefont {Frederico}, \citenamefont {de~Paula},\ and\ \citenamefont
  {Santos}}]{Chaves14}%
  \BibitemOpen
  \bibfield  {author} {\bibinfo {author} {\bibfnamefont {A.~J.}\ \bibnamefont
  {Chaves}}, \bibinfo {author} {\bibfnamefont {T.}~\bibnamefont {Frederico}},
  \bibinfo {author} {\bibfnamefont {W.}~\bibnamefont {de~Paula}}, \ and\
  \bibinfo {author} {\bibfnamefont {M.~C.}\ \bibnamefont {Santos}},\ }\href
  {\doibase 10.1088/0953-8984/26/18/185301} {\bibfield  {journal} {\bibinfo
  {journal} {J. Phys.: Condens. Matter}\ }\textbf {\bibinfo {volume} {26}},\
  \bibinfo {pages} {1853301} (\bibinfo {year} {2014})}\BibitemShut {NoStop}%
\bibitem [{\citenamefont {Ramon}(1989)}]{Ramond1990}%
  \BibitemOpen
  \bibfield  {author} {\bibinfo {author} {\bibfnamefont {P.}~\bibnamefont
  {Ramon}},\ }\href@noop {} {\emph {\bibinfo {title} {{Field Theory: A Modern
  Primer}}}}\ (\bibinfo  {publisher} {Addison-Wesley},\ \bibinfo {address} {New
  York},\ \bibinfo {year} {1989})\BibitemShut {NoStop}%
\bibitem [{\citenamefont {de~Sabbata}\ and\ \citenamefont
  {Gasperini}(1985)}]{Sabbata1985}%
  \BibitemOpen
  \bibfield  {author} {\bibinfo {author} {\bibfnamefont {V.}~\bibnamefont
  {de~Sabbata}}\ and\ \bibinfo {author} {\bibfnamefont {M.}~\bibnamefont
  {Gasperini}},\ }\href@noop {} {\emph {\bibinfo {title} {{Introduction to
  Gravitation}}}}\ (\bibinfo  {publisher} {World Scientific},\ \bibinfo
  {address} {Singapore},\ \bibinfo {year} {1985})\BibitemShut {NoStop}%
\bibitem [{\citenamefont {de~Juan}\ \emph {et~al.}(2009)\citenamefont
  {de~Juan}, \citenamefont {Cortijo},\ and\ \citenamefont
  {Vozmediano}}]{Vozmediano09}%
  \BibitemOpen
  \bibfield  {author} {\bibinfo {author} {\bibfnamefont {F.}~\bibnamefont
  {de~Juan}}, \bibinfo {author} {\bibfnamefont {A.}~\bibnamefont {Cortijo}}, \
  and\ \bibinfo {author} {\bibfnamefont {M.~A.~H.}\ \bibnamefont
  {Vozmediano}},\ }\href {\doibase 10.1016/j.nuclphysb.2009.11.012} {\bibfield
  {journal} {\bibinfo  {journal} {Nucl. Phys. B}\ }\textbf {\bibinfo {volume}
  {828}},\ \bibinfo {pages} {625} (\bibinfo {year} {2009})}\BibitemShut
  {NoStop}%
\bibitem [{\citenamefont {Ishigami}\ \emph {et~al.}(2007)\citenamefont
  {Ishigami}, \citenamefont {Chen}, \citenamefont {Cullen}, \citenamefont
  {Fuhrer},\ and\ \citenamefont {Williams}}]{Ishigami07}%
  \BibitemOpen
  \bibfield  {author} {\bibinfo {author} {\bibfnamefont {M.}~\bibnamefont
  {Ishigami}}, \bibinfo {author} {\bibfnamefont {J.~H.}\ \bibnamefont {Chen}},
  \bibinfo {author} {\bibfnamefont {W.~G.}\ \bibnamefont {Cullen}}, \bibinfo
  {author} {\bibfnamefont {M.~S.}\ \bibnamefont {Fuhrer}}, \ and\ \bibinfo
  {author} {\bibfnamefont {E.~D.}\ \bibnamefont {Williams}},\ }\href {\doibase
  10.1021/nl070613a} {\bibfield  {journal} {\bibinfo  {journal} {Nano Lett.}\
  }\textbf {\bibinfo {volume} {7}},\ \bibinfo {pages} {1643} (\bibinfo {year}
  {2007})}\BibitemShut {NoStop}%
\bibitem [{\citenamefont {Geringer}\ \emph {et~al.}(2009)\citenamefont
  {Geringer}, \citenamefont {Liebmann}, \citenamefont {Echtermeyer},
  \citenamefont {Runte}, \citenamefont {Schmidt}, \citenamefont {R\"{u}ckamp},
  \citenamefont {Lemme},\ and\ \citenamefont {Morgenstern}}]{Geringer09}%
  \BibitemOpen
  \bibfield  {author} {\bibinfo {author} {\bibfnamefont {V.}~\bibnamefont
  {Geringer}}, \bibinfo {author} {\bibfnamefont {M.}~\bibnamefont {Liebmann}},
  \bibinfo {author} {\bibfnamefont {T.}~\bibnamefont {Echtermeyer}}, \bibinfo
  {author} {\bibfnamefont {S.}~\bibnamefont {Runte}}, \bibinfo {author}
  {\bibfnamefont {M.}~\bibnamefont {Schmidt}}, \bibinfo {author} {\bibfnamefont
  {R.}~\bibnamefont {R\"{u}ckamp}}, \bibinfo {author} {\bibfnamefont {M.~C.}\
  \bibnamefont {Lemme}}, \ and\ \bibinfo {author} {\bibfnamefont
  {M.}~\bibnamefont {Morgenstern}},\ }\href {\doibase
  10.1103/PhysRevLett.102.076102} {\bibfield  {journal} {\bibinfo  {journal}
  {Phys. Rev. Lett.}\ }\textbf {\bibinfo {volume} {102}},\ \bibinfo {pages}
  {076102} (\bibinfo {year} {2009})}\BibitemShut {NoStop}%
\bibitem [{\citenamefont {Atalaya}\ \emph {et~al.}(2008)\citenamefont
  {Atalaya}, \citenamefont {Isacsson},\ and\ \citenamefont
  {Kinaret}}]{Atalaya08}%
  \BibitemOpen
  \bibfield  {author} {\bibinfo {author} {\bibfnamefont {J.}~\bibnamefont
  {Atalaya}}, \bibinfo {author} {\bibfnamefont {A.}~\bibnamefont {Isacsson}}, \
  and\ \bibinfo {author} {\bibfnamefont {J.~M.}\ \bibnamefont {Kinaret}},\
  }\href {\doibase 10.1021/nl801733d} {\bibfield  {journal} {\bibinfo
  {journal} {Nano Lett.}\ }\textbf {\bibinfo {volume} {8}},\ \bibinfo {pages}
  {4196} (\bibinfo {year} {2008})}\BibitemShut {NoStop}%
\bibitem [{\citenamefont {Yang}(2014)}]{Yang2014}%
  \BibitemOpen
  \bibfield  {author} {\bibinfo {author} {\bibfnamefont {B.}~\bibnamefont
  {Yang}},\ }\href {\doibase 10.1103/PhysRevB.91.241403} {\bibfield  {journal}
  {\bibinfo  {journal} {Phys. Rev. B}\ }\textbf {\bibinfo {volume} {91}},\
  \bibinfo {pages} {241403} (\bibinfo {year} {2015})}\BibitemShut {NoStop}%
\end{thebibliography}
%

\end{document}